\documentclass[sigconf]{acmart}
\usepackage{mathtools}
\usepackage{natbib}
\usepackage{multicol, multirow}
\usepackage{threeparttable}
\usepackage{booktabs}
\usepackage{amsmath}
\usepackage{siunitx}

\usepackage{graphicx}
\usepackage{subcaption}
\usepackage{algorithm}
\usepackage{algpseudocode}
\usepackage{ulem}
\usepackage{caption}
\usepackage{bbding}
\usepackage{enumitem}
\usepackage{dutchcal}



\begin{document}
\title{Exploring Test-time Scaling via Prediction Merging on Large-Scale Recommendation}

\author{Fuyuan Lyu}
\authornote{Equal Contribution} \affiliation{ \institution{MBZUAI \& McGill University \& \\MILA - Quebec AI Institute} \city{Montreal} \country{Canada} }
\email{fuyuan.lyu@mail.mcgill.ca}
\orcid{0000-0001-9345-1828}

\author{Zhentai Chen}
\authornotemark[1] \affiliation{ \institution{School of Artificial Intelligence \\ Shenzhen Technology University} \city{Shenzhen} \country{China} }
\email{zenithaitran@gmail.com}
\orcid{0009-0008-4824-1889}

\author{Jingyan Jiang}
\affiliation{ \institution{School of Artificial Intelligence \\ Shenzhen Technology University} \city{Shenzhen} \country{China} }
\email{jiangjingyan@sztu.edu.cn}
\orcid{0000-0003-4897-2645}

\author{Lingjie Li}
\affiliation{ \institution{School of Artificial Intelligence \\ Shenzhen Technology University} \city{Shenzhen} \country{China} }
\email{lilingjie2017@email.szu.edu.cn}

\author{Xing Tang}
\authornote{Corresponding authors} \affiliation{ \institution{School of Artificial Intelligence \\ Shenzhen Technology University} \city{Shenzhen} \country{China} }
\email{xing.tang@hotmail.com}
\orcid{0000-0003-4360-0754}

\author{Xiuqiang He}
\affiliation{ \institution{School of Artificial Intelligence \\ Shenzhen Technology University} \city{Shenzhen} \country{China} }
\email{he.xiuqiang@gmail.com}
\orcid{0000-0002-4115-8205}

\author{Xue Liu}
\affiliation{ \institution{MBZUAI \& McGill University} \city{Abu Dhabi} \country{UAE} }
\email{steve.liu@mbzuai.ac.ae}
\orcid{0000-0001-5252-3442}

\newcommand{\fuyuan}[1]{{\bf \color{red} [Fuyuan says ``#1'']}}
\newcommand{\Zenith}[1]{{\bf \color{blue} [Zenith says ``#1'']}}

\begin{abstract}
Inspired by the success of language models (LM), scaling up deep learning recommendation systems (DLRS) has become a recent trend in the community.
All previous methods tend to scale up the model parameters during training time.
However, how to efficiently utilize and scale up computational resources during test time remains underexplored, which can prove to be a scaling-efficient approach and bring orthogonal improvements in LM domains.
The key point in applying test-time scaling to DLRS lies in effectively generating diverse yet meaningful outputs for the same instance.
We propose two ways: One is to explore the heterogeneity of different model architectures.
The other is to utilize the randomness of model initialization under a homogeneous architecture.
The evaluation is conducted across eight models, including both classic and SOTA models, on three benchmarks. 
Sufficient evidence proves the effectiveness of both solutions.
We further prove that under the same inference budget, test-time scaling can outperform parameter scaling.
Our test-time scaling can also be seamlessly accelerated with the increase in parallel servers when deployed online, without affecting the inference time on the user side.
Code is available here\footnote{https://github.com/aTitye/TTS4CTR}.
\end{abstract}

\maketitle


\section{Introduction}
\label{sec:intro}
Deep learning-based recommendation systems (DLRS) are deemed essential for multiple online services these days~\cite{DLRM}. Modern DLRS takes multiple continuous dense features, such as date, and categorical sparse features, such as user clicked history. Categorical features are transformed into dense embedding representations throughout trainable embedding components. These dense embeddings are then fed into an interaction \& prediction component, designed to explicitly capture the intricate interactions between features in an efficient manner~\cite{IPNN,PIN}.

Driven by the advancements in scaling law in the Language Model (LM) domain~\cite{ScalingLaw}, scaling up the DLRS to fully leverage large volumes of online data remains an urgent need. 
Currently, the scaling on large-scale recommendation systems expands over two dimensions: (i) \textbf{width scaling}, which expands the number of interaction modules in a parallel manner. To avoid collisions and achieve better representation, each module tends to correspond with a unique and independently trainable embedding table~\cite{CETNet,D-MoE}. Such scaling methods appear similar to Mixture-of-Expert~\cite{MMoE,MoE}, but feature multiple embedding tables~\cite{MultiEmbed}. and (ii) \textbf{depth scaling}, which scales up the depth of the feature interaction module by designing a unified architecture that can be seamlessly stacked together~\cite{Wukong,RankMixer,SUAN}. However, both lines of work aim to scale up a single model by increasing the number of trainable parameters.

\begin{figure}[!htbp]
\centering
\includegraphics[width=0.48\textwidth]{figs/intro.png}
\caption{Comparison between Parameter Scaling and Test-time Scaling via Prediction Merging. The latter can benefit from additional improvement, superior stability, higher efficiency, and easy deployment.}
\Description{Comparison between Parameter Scaling and Test-time Scaling via Prediction Merging.}
\label{fig:intro}
\end{figure}

Recently, the test-time scaling techniques have gathered increasing attention~\cite{optimal-compute,s1}. The core idea of test-time scaling in the LM domain is to \textit{deliberately increase computation at test time in the aim of increasing the quality of results}~\cite{optimal-compute}. Test-time scaling has achieved \textbf{significant}, \textbf{resource efficient}, and \textbf{orthogonal} improvement over the parameter scaling solutions. Test-time scaling can also provide superior stability and ease of deployment in the DLRS domain, as elaborated in later Sections. Figure~\ref{fig:intro} visually explains the general idea and benefit of test-time scaling in DLRS. Yet, such an important idea has little exploration.

Although being a promising direction, the concept of test-time scaling cannot be effortlessly extended to DLRS. LMs can, by nature, generate diverse outputs, given the randomness in the factors such as the prompt or the decoding stage. However, for recommendations, especially those on a large scale, uniform input structures are preferred due to the requirement for large-scale online serving. Besides, the recommendation models tend to be trained under a unified data distribution, namely, all the features deemed useful by experts or algorithms~\cite{AutoField,OptFS} are input into the model to achieve the best performance possible. Hence, how to obtain \textbf{diverse yet meaningful} outputs during test-time over large-scale recommendation models directly affects the utility of test-time scaling. 

Here, we propose two approaches for obtaining diverse outputs from the same input. One approach is to leverage the heterogeneity among different recommendation models. Recent works have consistently proven the effectiveness of utilizing different recommendation architectures in capturing diverse interactions and correlations~\cite{D-MoE}. The other approach involves exploring randomness within a homogeneous architecture. Specifically, as initialization plays a crucial role in model training~\cite{lth}, we deliberately train models with different random seeds to encourage diversity in the logits spaces. Note that these two approaches can be combined.
To efficiently fuse all the generated outputs, we propose to directly fuse them in the prediction space, namely, prediction merging.
 
We conduct extensive evaluations of eight recommendation models, both classic and SOTA, and their combinations. The evaluation is conducted across three benchmarks: Criteo, Avazu, and KDD12, with significant improvements observed in all cases. We further demonstrate the efficiency, the stability, and the orthogonality of test-time scaling. Further investigation reveals that diversity among predictions is a key factor in driving performance increases. Importantly, as test-time scaling via prediction merging is by nature parallel, it can be easily deployed across different servers without a significant reduction in inference speed.
Our contribution can be summarized as follows:
\begin{itemize}[topsep=0pt,noitemsep,nolistsep,leftmargin=*]
    \item We formally introduce an additional scaling dimension in DLRS: test-time scaling, which is orthogonal to the current parameter scaling approaches.
    \item To effectively obtain diverse prediction logits via test-time, we propose two solutions of scaling under prediction merging for both homogeneous and hierarchical architectures. 
    \item Extensive evaluation over three large-scale benchmarks, eight recommendation models, and their combinations, validates the effectiveness, efficiency, and orthogonality of test-time scaling via prediction merging.
\end{itemize}

\section{Methodology}
\label{sec:method}

\begin{figure*}[!htbp]
\centering
\includegraphics[width=0.85\textwidth]{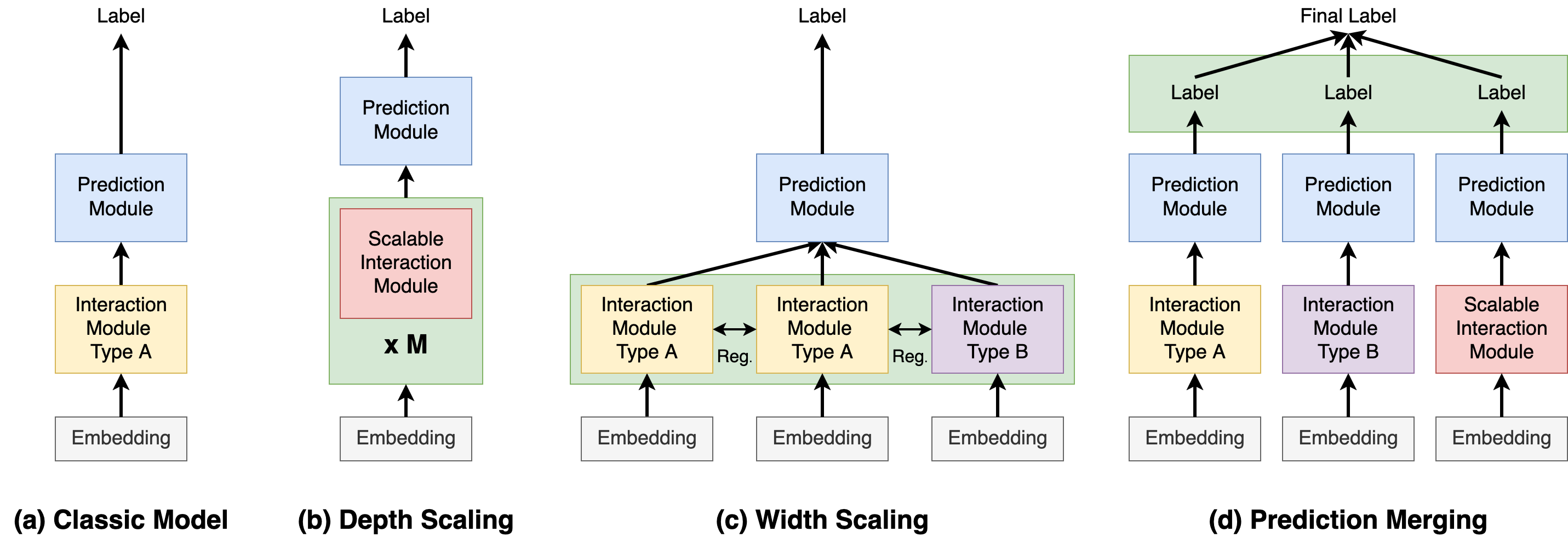}
\caption{Visualization of different structures. (a) Classic Model design with Embedding, Interaction, and Prediction Module. (b) Depth Scabling Model design, which scales the depth of the interaction module and implicitly fuses information. (c) The Width Scaling Model design increases the number of interaction modules and the associated embedding table in parallel. It applies regularization among the interaction modules to encourage diversity and prevent overlapping. (d) Prediction Merging fuses the information directly in the logit space. For all structures, the interaction module and prediction module can have the same or different architectures. {\color{green}Green} box indicates the information fusion process.}
\Description{Visualization of different structures.}
\label{fig:method}
\end{figure*}

\subsection{Problem Definition}
Large-scale Recommendation tends to formulate the problem as a binary classification. Specifically, it aims to predict the probability of a user interacting with given items. It can be defined as $\{\mathcal{X}, \mathcal{Y}\}$, with cardinality $|\{\mathcal{X}, \mathcal{Y}\}| = N$, where $\mathcal{X}$ is a multi-categorical feature set, including user, item, or contextual features, and $\mathcal{Y} \in \{0,1\}^N$ is a label set indicating whether the user interacted with the given item. 
Each feature $x_i$ belongs to a feature field $\mathbf{x}_i$ with cardinality $D_i$, where $D_i$ is the number of unique features in the feature field. Each feature field $\mathbf{x}_i$ is associated with an embedding table $\mathbf{E} \in \mathbb{R}^{D_i \times d}$, where $d$ is the pre-defined embedding dimension size. This can be formulated as:
\begin{equation}
    \mathbf{e}_i = \mathbf{E}(x_i).
\end{equation}
All these embeddings are concatenated as $\mathbf{e} = [\mathbf{e}_1, \cdots, \mathbf{e}_n]$, where $n$ indicates the number of feature fields. 
After obtaining all the embeddings, an interaction layer is applied to the embeddings, aiming to extract unique interactions. This can be formulated as:
\begin{equation}
    \mathbf{h} = I(\mathbf{e}).
\end{equation}
A prediction module is further concatenated to learn a mapping between the interaction space $I$ and prediction label $\mathcal{Y}$, which can be formulated as: $\mathcal{F}: I \rightarrow \mathcal{Y}$.
Binary cross-entropy (BCE) loss is commonly chosen as the objective function. For each instance $(x, y)\in \{\mathcal{X}, \mathcal{Y}\}$, the model can make the prediction as: $\hat{y} = \mathcal{F}(\mathbf{h})$. Hence, the final loss is formulated as follows:
\begin{equation} \label{eq:loss_pred}
\mathcal{L} = \mathcal{L}_{\text{pred}} = - \frac{1}{N} \sum_{x,y \in \{\mathcal{X}, \mathcal{Y}\}}^N y\log(\hat{y}) + (1 - y) \log(1-\hat{y}).
\end{equation}

\subsection{Parameter Scaling}
\label{sec:method_param}
In this section, we briefly categorize and formulate the two classes of parameter scaling. 

\subsubsection{Depth Scaling}
Depth scaling~\cite{Wukong,RankMixer,SUAN} methods focus on increasing the depth of the interaction layer. One of the key bottlenecks lies in designing a scalable interaction layer. Hence, various layers, specifically tailored to be scalable and seamlessly concatenated, have been proposed. From a high-level perspective, we can formulate such a design as proposing a series of interaction layers $\{I^{(i)} | 1 \leq i \leq M\}$, and each layer is stacked upon the former with the initial representation $\mathbf{h}^{(0)}$ being the embedding $\mathbf{e}$:
\begin{equation}
    \mathbf{h}^{(i)} = I^{(i)}(\mathbf{h}^{(i-1)}), \ \mathbf{h}^{(0)} = \mathbf{e}.
\end{equation}
Finally, the prediction module maps the final representation into the logits space, denoted as $\hat{y} = \mathcal{F}(\mathbf{h}^{(M)})$. The depth scaling models tend to be trained using similar objectives as Equation~\ref{eq:loss_pred}.

\subsubsection{Width Scaling}

Width Scaling~\cite{MultiEmbed,D-MoE,CETNet} methods, on the other hand, intend to increase the width of the interaction layer. It is based on the hypothesis that various interaction layers may have different expertise in capturing users' interest. Hence, concatenating them may yield a better representation of users' interests. 
Due to the embedding collapse phenomenon~\cite{MultiEmbed}, a trainable embedding table is associated with the corresponding interaction component for better scalability and capturing diverse behaviour patterns~\cite{MMoE,D-MoE}. Specifically, a model with $M$ sets of embedding tables is defined as:
\begin{equation}
\begin{aligned}
    \mathbf{e}^{(m)}_i &= \mathbf{E}^{(m)}(x_i), \ 1 \leq m \leq M \\
    \mathbf{e}^{(m)} &= [\mathbf{e}^{(m)}_1, \cdots, \mathbf{e}^{(m)}_n].
\end{aligned}
\end{equation}
After obtaining the $M$ sets of embedding $\{\mathbf{e}^{(m)} \ | \ 1 \leq m \leq M \}$, single or multiple experts, referring to different interaction modules, such as CIN~\cite{xDeepFM}, Cross Layer~\cite{DCN,DCNv2}, or Factorization Machine~\cite{FM,DeepFM}, may be utilized to capture the diverse behaviour patterns. Specificially, each interaction expert $I^{(m)}$ is assigned to the $m$ embedding $\mathbf{e}^{(m)}$ and aims to capture diverse interactions, formulated as:
\begin{equation}
    \mathbf{h}^{(m)} = I^{(m)}(\mathbf{e}^{(m)}).
\end{equation}
Here, each different interaction modules are averaged to obtain the final interactions
\begin{equation}
    \mathbf{h} = \frac{1}{M}\sum_{m=1}^M h^{(m)},
\end{equation}
and the final prediction can be obtained as:$\hat{y} = \mathcal{F}(\mathbf{h})$.
Additionally, to avoid the similarity among interaction representations, various diversity regularization terms~\cite{D-MoE,CETNet} have been proposed. This can be followed as:
\begin{equation}
    \mathcal{L} = \mathcal{L}_{\text{pred}} + \lambda \mathcal{L}_{\text{reg}}(\{\mathbf{h}^{(m)} \ | \ 1 \leq m \leq M\}), 
\end{equation}
with $\lambda$ indicating the strength of regularization and $\mathcal{L}_{\text{reg}}$ being the regularization term encouraging diversity among the interaction set $\{\mathbf{h}^{(m)} \ | \ 1 \leq m \leq M\}$. 

\subsection{Test-time Scaling via Prediction Merging}
\label{sec:method_pm}

The major difference between existing parameter scaling methods and prediction merging lies in two aspects. First, prediction merging encourages the generation of diverse outputs directly, whereas parameter scaling methods only generate a single prediction. Unlike existing models~\cite{D-MoE,CETNet,MultiEmbed,Wukong}, which operate on the interaction layer, prediction merging directly operates on the prediction logit space $\mathcal{\hat{Y}}$ without making assumptions about the architecture of the individual model. Second, the number of supervision signals for prediction merging is higher than that for parameter scaling. Prediction merging involves training models $\mathcal{G}^{(m)}: x \rightarrow y$ independently or in a grouped manner, whereas existing solutions train all experts jointly with only one supervision signal, incorporating regularization based on human-prior knowledge. 

Hence, under the prediction merging framework, multiple models, regardless of their architectures or prior assumptions, can be integrated seamlessly. Specifically, given a family of models $\{\mathcal{F}^{(m)} \ | \ 1 \leq m \leq M\}$. Hence, for each data instance $(x, y) \in \{\mathcal{X}, \mathcal{Y}\}$, model $\mathcal{F}^{(m)}$ make its prediction as
\begin{equation}
\hat{y}^{(m)} = \mathcal{G}^{(m)}(x).
\end{equation}
After obtaining the prediction sets $\{\hat{y}^{(m)} \ | \ 1 \leq m \leq M \}$, prediction merging directly averages all predictions and yields the final prediction as:
\begin{equation} \label{eq:average}
\tilde{y} = \frac{1}{M} \sum_{m=1}^M \hat{y}^{(m)}.
\end{equation}

Algorithm~\ref{alg:tts} describe the detailed solution in obtaining the model set $\{ \mathcal{G}^{(m)} \ | \ 1 \leq m \leq M \}$. Ablation study on the grouped training is provided in Appendix~\ref{appendix:hyperparam}.
\begin{algorithm}
	\caption{The Training Phase for Prediction Merging} 
    \label{alg:tts}
	\begin{algorithmic}[1]
		\Require training dataset $\{\mathcal{X}, \mathcal{Y}\}$, initial model set $\{ \mathcal{G}_0^{(m)} \ | \ 1 \leq m \leq M \}$
        \Ensure model set $\{ \mathcal{G}^{(m)} \ | \ 1 \leq m \leq M \}$
        
        \State Split initial model set into $K$ mutually excluded subset $\mathcal{S}_1, \cdots, \mathcal{S}_K$, with each containing at least one element.
        \State k = 0
        \While {k < K}
            \State k = k + 1
            \While{not converged}
                \State Sample a mini-batch from the training dataset $\{\mathcal{X}, \mathcal{Y}\}$
                \State Calculate the prediction $\tilde{y}_k$ with Equation~\ref{eq:average} given $\mathcal{S}_k$
                \State Tune models in $\mathcal{S}_k$ given the objective in Equation \ref{eq:loss_pred}
            \EndWhile
        \EndWhile 
	\end{algorithmic}
\end{algorithm}

\section{Experiment}
\label{sec:experiment}

In this section, we aim to evaluate and answer the following research questions:
\begin{itemize}[topsep=0pt,noitemsep,nolistsep,leftmargin=*]
    \item \textbf{RQ1}: Is test-time scaling effective for recommendation?
    \item \textbf{RQ2}: How does test-time scaling compare to depth scaling models?
    \item \textbf{RQ3}: How does test-time scaling compare to width scaling methods?
    \item \textbf{RQ4}: Is test-time scaling orthogonal to parameter scaling?
    \item \textbf{RQ5}: How does prediction merging compare to model merging?
    \item \textbf{RQ6}: How does test-time scaling compare to parameter scaling in terms of efficiency in the real world?
    \item \textbf{RQ7}: Why can test-time scaling be achieved via prediction merging?
\end{itemize}

\subsection{Experimental Setup}
\subsubsection{Datasets}
We conduct our experiments on three public real-world datasets. We describe all datasets and the pre-processing steps below. Statistics are provided in Appendix~\ref{appendix:statistic}.

\textbf{Criteo}~\cite{Criteo-dataset} dataset consists of ad click data over a week. It consists of 26 categorical feature fields and 13 numerical feature fields. We discretize each numeric value $x$ to $\lfloor\log^2(x)\rfloor$, if $x>2$; $x=1$ otherwise. We replace infrequent categorical features with a default "OOV" (i.e., out-of-vocabulary) token, with \textit{min\_count}=10.

\textbf{Avazu}~\cite{Avazu-dataset} dataset contains 10 days of click logs. It has 24 fields with categorical features. We remove the \textit{instance\_id} field and transform the \textit{timestamp} field into three new fields: \textit{hour}, \textit{weekday}, and \textit{is\_weekend}. We replace infrequent categorical features with the "OOV" token, with \textit{min\_count}=2.

\textbf{KDD12}~\cite{KDD12-dataset} dataset contains training instances derived from se-arch session logs. It has 11 categorical fields, and the click field is the number of times the user clicks the ad. We replace infrequent features with an "OOV" token, with \textit{min\_count}=10.

\subsubsection{Metrics}
Following the previous works~\cite{FM,DeepFM}, we use the common evaluation metrics for CTR prediction: \textbf{AUC} (Area Under ROC) and \textbf{Log loss} (cross-entropy). Note that $\mathbf{0.1 \%}$ improvement in AUC is considered significant~\cite{IPNN,DeepFM}.

\subsubsection{Backbones and Baselines}
We evaluate the test-time scaling and other scaling techniques on two types of backbone models: (i) classical DLRS models: FNN~\cite{FNN}, IPNN~\cite{IPNN}, DeepFM~\cite{DeepFM}, DCN~\cite{DCN}, DCNv2~\cite{DCNv2}, and FinalMLP~\cite{FinalMLP}, and (ii) depth scaling models: Wu-kong~\cite{Wukong} and RankMixer~\cite{RankMixer}. We further compare with two width scaling techniques orthogonal to backbone models, namely Multi-Embed (ME)~\cite{MultiEmbed}, D-MoE~\cite{D-MoE}. 

\subsection{RQ1: Main Results} \label{sec:experiment_main}

In this section, we evaluate the scalability of prediction merging in two situations: (i) homogenous architectures and (ii) heterogeneous architectures.

\begin{table*}[!htbp]
    \centering
    \caption{Scalability on Homogenous Architecture.}
    \label{table:main_homo}
    \vspace{-10pt}
    \resizebox{0.98\textwidth}{!}{{}
    \begin{tabular}{c|c|cc|cc|cc|cc|cc|cc|cc|cc}
        \hline
            & \multirow{2}{*}{Method} & \multicolumn{2}{c}{FNN} & \multicolumn{2}{|c}{IPNN} & \multicolumn{2}{|c}{DeepFM} & \multicolumn{2}{|c}{DCN} & \multicolumn{2}{|c}{DCNv2} & \multicolumn{2}{|c}{FinalMLP} & \multicolumn{2}{|c}{Wukong} & \multicolumn{2}{|c}{RankMixer} \\
        \cline{3-18}     &  & AUC$\uparrow$     & Logloss$\downarrow$ & AUC$\uparrow$  & Logloss$\downarrow$& AUC$\uparrow$ & Logloss$\downarrow$     & AUC$\uparrow$  & Logloss$\downarrow$     & AUC$\uparrow$ & Logloss$\downarrow$ & AUC$\uparrow$ & Logloss$\downarrow$ & AUC$\uparrow$ & Logloss$\downarrow$ & AUC$\uparrow$ & Logloss$\downarrow$ \\
        \hline
        \multirow{5}{*}{\rotatebox{90}{Avazu}}  
            & Backbone   & 0.7885 & 0.3764 & 0.7914 & 0.3743 & 0.7869 & 0.3757 & 0.7882 & 0.3766 & 0.7946 & 0.3711 & 0.7858 & 0.3777 & 0.7954 & 0.3704 & 0.7952 & 0.3706 \\
            & 2$\times$  & 0.7895$^*$ & 0.3757 & 0.7926$^*$ & 0.3733 & 0.7871 & 0.3755 & 0.7896$^*$ & 0.3754 & 0.7957$^*$ & 0.3704 & 0.7882$^*$ & 0.3758 & 0.7972$^*$ & 0.3694 & 0.7963$^*$ & 0.3699 \\
            & 4$\times$  & 0.7903$^*$ & 0.3748 & 0.7934$^*$ & 0.3728 & 0.7873 & 0.3754 & 0.7903$^*$ & 0.3749 & 0.7969$^*$ & 0.3698 & 0.7892$^*$ & 0.3744 & 0.7986$^*$ & 0.3685 & 0.7972$^*$ & 0.3693 \\
            & 8$\times$  & 0.7906$^*$ & 0.3747 & 0.7938$^*$ & 0.3725 & 0.7873 & 0.3754 & 0.7906$^*$ & 0.3747 & 0.7979$^*$ & 0.3692 & 0.7898$^*$ & 0.3741 & 0.7994$^*$ & 0.3681 & 0.7981$^*$ & 0.3689 \\
            & 16$\times$ & 0.7906$^*$ & 0.3746 & 0.7940$^*$ & 0.3723 & 0.7873 & 0.3754 & 0.7908$^*$ & 0.3745 & 0.7989$^*$ & 0.3685 & 0.7904$^*$ & 0.3737 & 0.8001$^*$ & 0.3677 & 0.7988$^*$ & 0.3684 \\
        \hline
        \multirow{5}{*}{\rotatebox{90}{Criteo}} 
            & Backbone   & 0.8103 & 0.4411 & 0.8109 & 0.4405 & 0.8102 & 0.4413 & 0.8103 & 0.4412 & 0.8116 & 0.4398 & 0.8105 & 0.4410 & 0.8103 & 0.4411 & 0.8115 & 0.4400 \\
            & 2$\times$  & 0.8110 & 0.4405 & 0.8117 & 0.4397 & 0.8112 & 0.4404 & 0.8110 & 0.4405 & 0.8124 & 0.4391 & 0.8112 & 0.4404 & 0.8121$^*$ & 0.4395 & 0.8124$^*$ & 0.4392 \\
            & 4$\times$  & 0.8113$^*$ & 0.4402 & 0.8122$^*$ & 0.4393 & 0.8116$^*$ & 0.4400 & 0.8114$^*$ & 0.4401 & 0.8129$^*$ & 0.4386 & 0.8116$^*$ & 0.4400 & 0.8131$^*$ & 0.4387 & 0.8133$^*$ & 0.4383 \\
            & 8$\times$  & 0.8115$^*$ & 0.4400 & 0.8123$^*$ & 0.4391 & 0.8118$^*$ & 0.4399 & 0.8115$^*$ & 0.4400 & 0.8131$^*$ & 0.4384 & 0.8117$^*$ & 0.4398 & 0.8135$^*$ & 0.4383 & 0.8139$^*$ & 0.4378 \\
            & 16$\times$ & 0.8115$^*$ & 0.4399 & 0.8124$^*$ & 0.4390 & 0.8119$^*$ & 0.4397 & 0.8116$^*$ & 0.4399 & 0.8132$^*$ & 0.4384 & 0.8118$^*$ & 0.4398 & 0.8137$^*$ & 0.4381 & 0.8142$^*$ & 0.4375 \\
        \hline
        \multirow{5}{*}{\rotatebox{90}{KDD12}}  
            & Backbone   & 0.8108 & 0.1500 & 0.8118 & 0.1498 & 0.8086 & 0.1505 & 0.8106 & 0.1501 & 0.8114 & 0.1499 & 0.8111 & 0.1499 & 0.8117 & 0.1498 & 0.8124 & 0.1496 \\
            & 2$\times$  & 0.8114 & 0.1499 & 0.8127$^*$ & 0.1496 & 0.8099$^*$ & 0.1502 & 0.8113 & 0.1499 & 0.8119 & 0.1497 & 0.8116 & 0.1498 & 0.8128$^*$ & 0.1496 & 0.8133$^*$ & 0.1494 \\
            & 4$\times$  & 0.8116 & 0.1498 & 0.8131$^*$ & 0.1495 & 0.8106$^*$ & 0.1500 & 0.8115$^*$ & 0.1499 & 0.8122$^*$ & 0.1497 & 0.8120$^*$ & 0.1497 & 0.8133$^*$ & 0.1495 & 0.8139$^*$ & 0.1493 \\
            & 8$\times$  & 0.8118$^*$ & 0.1498 & 0.8133$^*$ & 0.1494 & 0.8110$^*$ & 0.1499 & 0.8117$^*$ & 0.1498 & 0.8123$^*$ & 0.1496 & 0.8122$^*$ & 0.1497 & 0.8135$^*$ & 0.1494 & 0.8144$^*$ & 0.1492 \\
            & 16$\times$ & 0.8119$^*$ & 0.1498 & 0.8134$^*$ & 0.1494 & 0.8112$^*$ & 0.1499 & 0.8118$^*$ & 0.1498 & 0.8124$^*$ & 0.1496 & 0.8122$^*$ & 0.1497 & 0.8137$^*$ & 0.1494 & 0.8148$^*$ & 0.1491 \\
        \hline
    \end{tabular}
    }

    \begin{tablenotes}
        \footnotesize \item[1] Here $*$ denotes statistically significant improvement
        (measured by a two-sided t-test with p-value $<0.05$) over the backbone in terms of AUC.
        \vspace{-10pt}
    \end{tablenotes}
\end{table*}

\subsubsection{Homogenous Architectures} \label{sec:experiment_main_homo}
For homogeneous architectures, we scale up the same architecture via initialization randomness, achieving up to 16$\times$ scaling. Based on Table~\ref{table:main_homo}, we can make the following observations.
Firstly, we can observe that the performance, as measured by both AUC and Logloss, consistently improves as the number of experts scales. Significant improvements are observed in most cases.
Secondly, on certain models, significant improvements can be observed when merging merely two models, hinting at the efficiency of test-time scaling on homogeneous architectures.
Finally, we can observe that depth scaling architectures, namely Wukog and Rankmixer, exhibit better scalability than others. Specifically, as the number of models increases, Wukong and Rankmixer are easier to exhibit significant improvements.
Note that we further scale selected architectures, namely DCNv2, RankMixer, and Wukong, by up to 64$\times$ in Appendix~\ref{appendix:extended}. The observation is consistent with the above conclusions.

\begin{table}[!htbp]
\renewcommand\arraystretch{1.2}
\centering
\caption{Scalability on Heterogeneous Architecture.}
\label{table:main_hetero}
\vspace{-10pt}
\resizebox{0.48\textwidth}{!}{
\begin{tabular}{c|c|cc|cc|cc}
\hline 
    & \multirow{2}{*}{Method} & \multicolumn{2}{c|}{Wukong+Rankmixer} & \multicolumn{2}{c|}{Rankmixer+DCNv2} & \multicolumn{2}{c}{Wukong+DCNv2} \\
\cline{3-8} 
    & & AUC$\uparrow$ & Logloss$\downarrow$ & AUC$\uparrow$ & Logloss$\downarrow$ & AUC$\uparrow$ & Logloss$\downarrow$\\
\hline
\multirow{7}{*}{\rotatebox{90}{Avazu}}  
    & 1$\times$  & 0.7982 & 0.3687 & 0.7956 & 0.3702 & 0.7970 & 0.3694 \\
    & 2$\times$  & 0.7995$^*$ & 0.3679 & 0.7963 & 0.3699 & 0.7975 & 0.3691 \\
    & 4$\times$  & 0.8008$^*$ & 0.3671 & 0.7964 & 0.3698 & 0.7979$^*$ & 0.3689 \\
    & 8$\times$  & 0.8017$^*$ & 0.3666 & 0.7965 & 0.3698 & 0.7980$^*$ & 0.3688 \\
    & 16$\times$ & 0.8026$^*$ & 0.3660 & 0.7966$^*$ & 0.3697 & 0.7981$^*$ & 0.3688 \\
    & 32$\times$ & 0.8030$^*$ & 0.3657 & 0.7966$^*$ & 0.3697 & 0.7982$^*$ & 0.3687 \\
    
\hline
\multirow{7}{*}{\rotatebox{90}{Criteo}} 
    & 1$\times$  & 0.8129 & 0.4387 & 0.8128 & 0.4388 & 0.8129 & 0.4387 \\
    & 2$\times$  & 0.8138$^*$ & 0.4379 & 0.8137$^*$ & 0.4379 & 0.8136 & 0.4380 \\
    & 4$\times$  & 0.8141$^*$ & 0.4376 & 0.8139$^*$ & 0.4377 & 0.8140$^*$ & 0.4377 \\
    & 8$\times$  & 0.8144$^*$ & 0.4374 & 0.8141$^*$ & 0.4376 & 0.8141$^*$ & 0.4376 \\
    & 16$\times$ & 0.8145$^*$ & 0.4373 & 0.8142$^*$ & 0.4375 & 0.8142$^*$ & 0.4375 \\
    & 32$\times$ & 0.8145$^*$ & 0.4373 & 0.8142$^*$ & 0.4374 & 0.8143$^*$ & 0.4375 \\
    
\hline
\multirow{7}{*}{\rotatebox{90}{KDD12}}  
    & 1$\times$  & 0.8133 & 0.1495 & 0.8128 & 0.1495 & 0.8127 & 0.1496 \\
    & 2$\times$  & 0.8139 & 0.1493 & 0.8131 & 0.1494 & 0.8132 & 0.1495 \\
    & 4$\times$  & 0.8145$^*$ & 0.1491 & 0.8132 & 0.1494 & 0.8134 & 0.1494 \\
    & 8$\times$  & 0.8150$^*$ & 0.1490 & 0.8133 & 0.1494 & 0.8135 & 0.1494 \\
    & 16$\times$ & 0.8153$^*$ & 0.1489 & 0.8134 & 0.1494 & 0.8136$^*$ & 0.1494 \\
    & 32$\times$ & 0.8155$^*$ & 0.1489 & 0.8134 & 0.1494 & 0.8136$^*$ & 0.1494 \\
    
\hline
\end{tabular}}

\begin{tablenotes}
    \footnotesize \item[1] \noindent $^*$ denotes significant improvement between the current scaling and the base model on AUC, measured by a two-tailed t-test with $p$-value smaller than 0.05.
    \vspace{-5pt}
\end{tablenotes}
\end{table}

\subsubsection{Heterogenous Architectures} \label{sec:experiment_main_hetero}
For heterogeneous architectures, we investigate three combinations: (1) Wukong + Rankmixer, (2) Rankmixer + DCNv2, and (3) Wukong + DCNv2. The result is shown in Table~\ref{table:main_hetero}. We can make the following observations.
First, similar to homogeneous architecture, prediction merging can exhibit similar scalability with respect to the number of models in heterogeneous architectures. The performance increases as the number of models increases.
Second, we observe that combining strong architectures can yield better performance. Specifically, Wukong + Rankmixer outperforms the other two combinations. This corresponds to the previous observation that Wukong and Rankmixer yield better scalability.

\subsection{RQ2: Comparison against Depth Scaling}
\begin{figure*}[!t]
    \centering
    \begin{subfigure}[b]{0.48\textwidth}
        \centering
        \includegraphics[width=\linewidth]{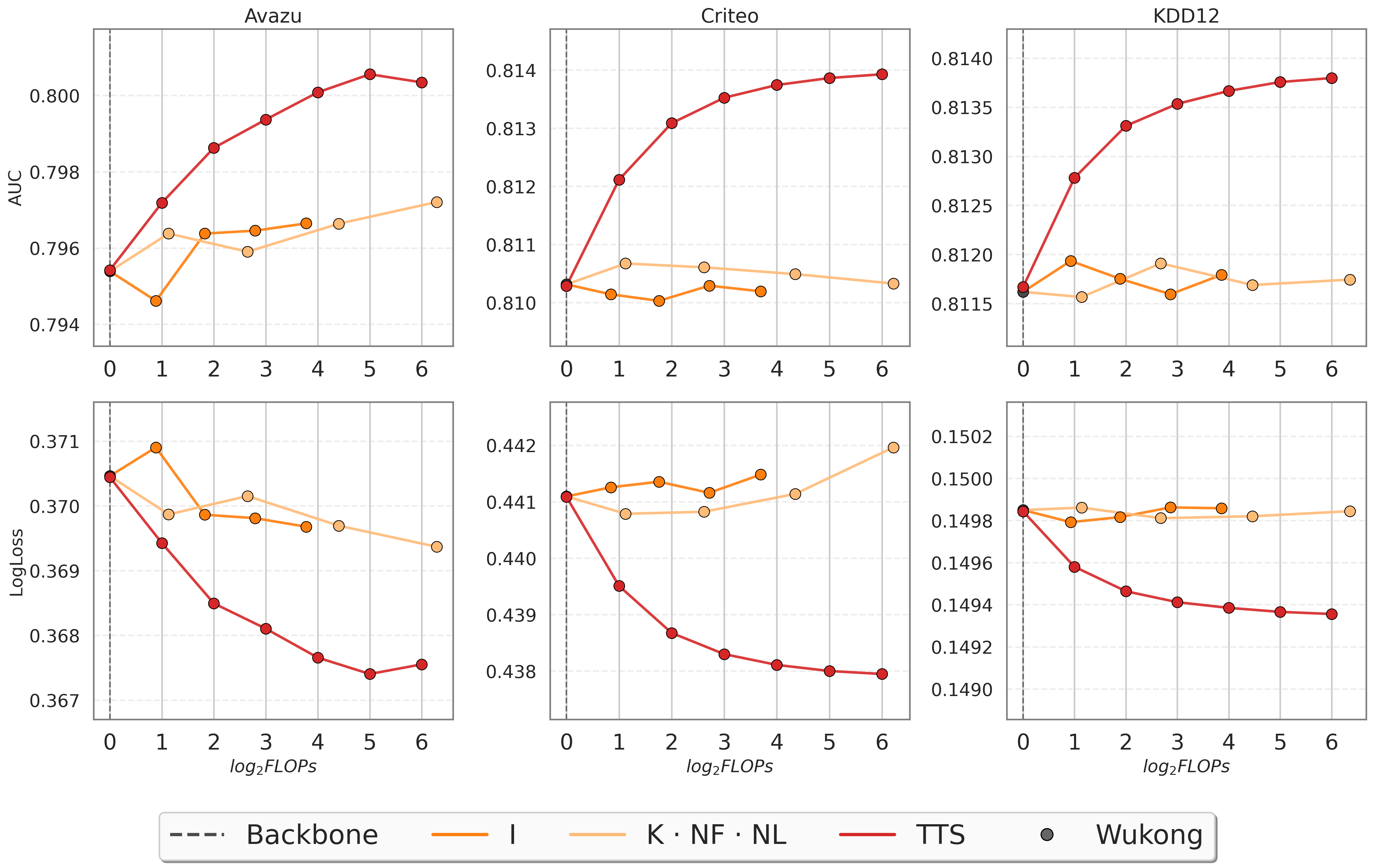}
        \caption{Wukong}
        \label{fig:wukong}
    \end{subfigure}
    \hfill
    \begin{subfigure}[b]{0.48\textwidth}
        \centering
        \includegraphics[width=\linewidth]{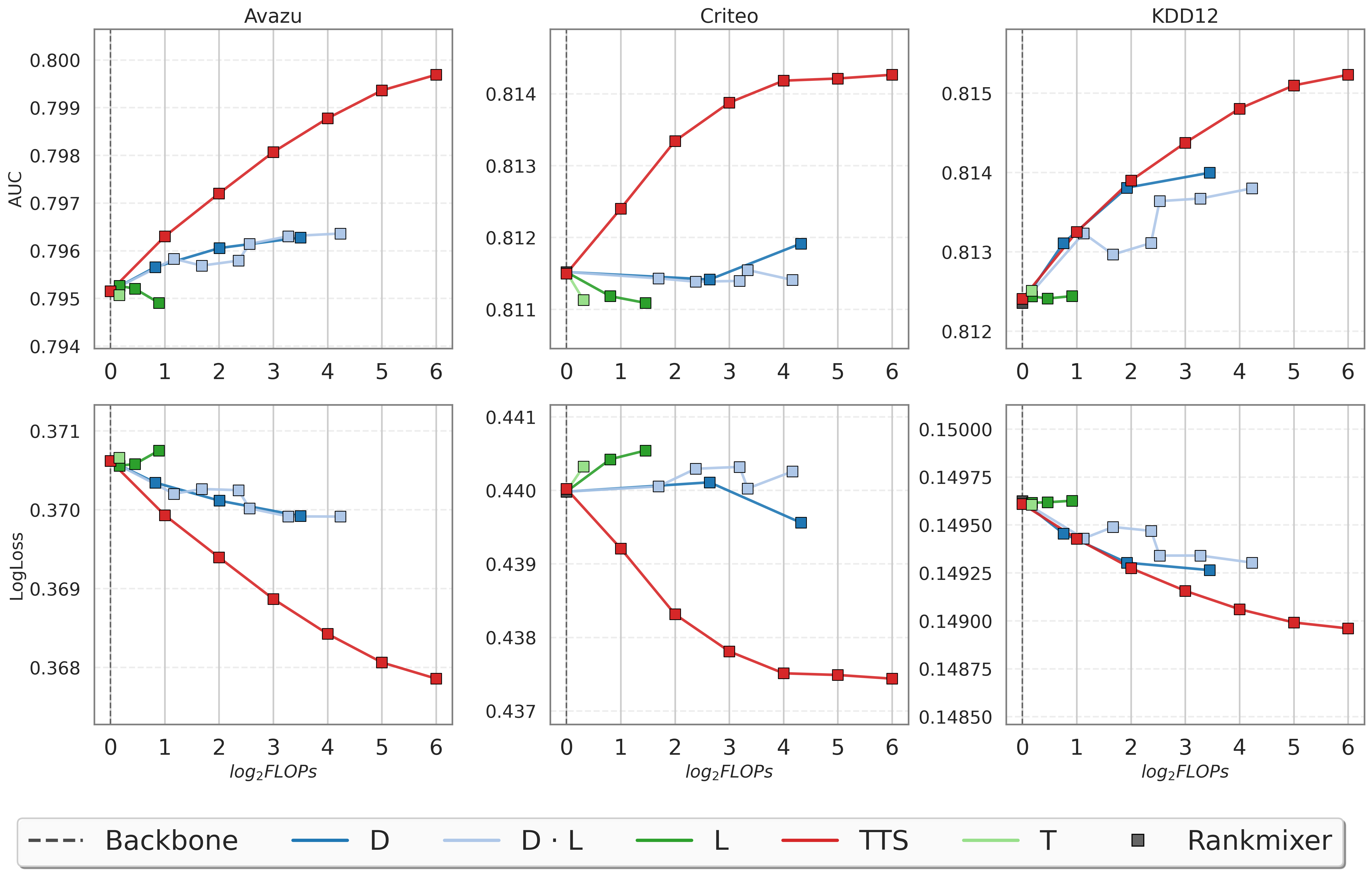}
        \caption{RankMixer}
        \label{fig:rankmixer}
    \end{subfigure}

    \begin{subfigure}[b]{0.48\textwidth}
        \centering
        \includegraphics[width=\linewidth]{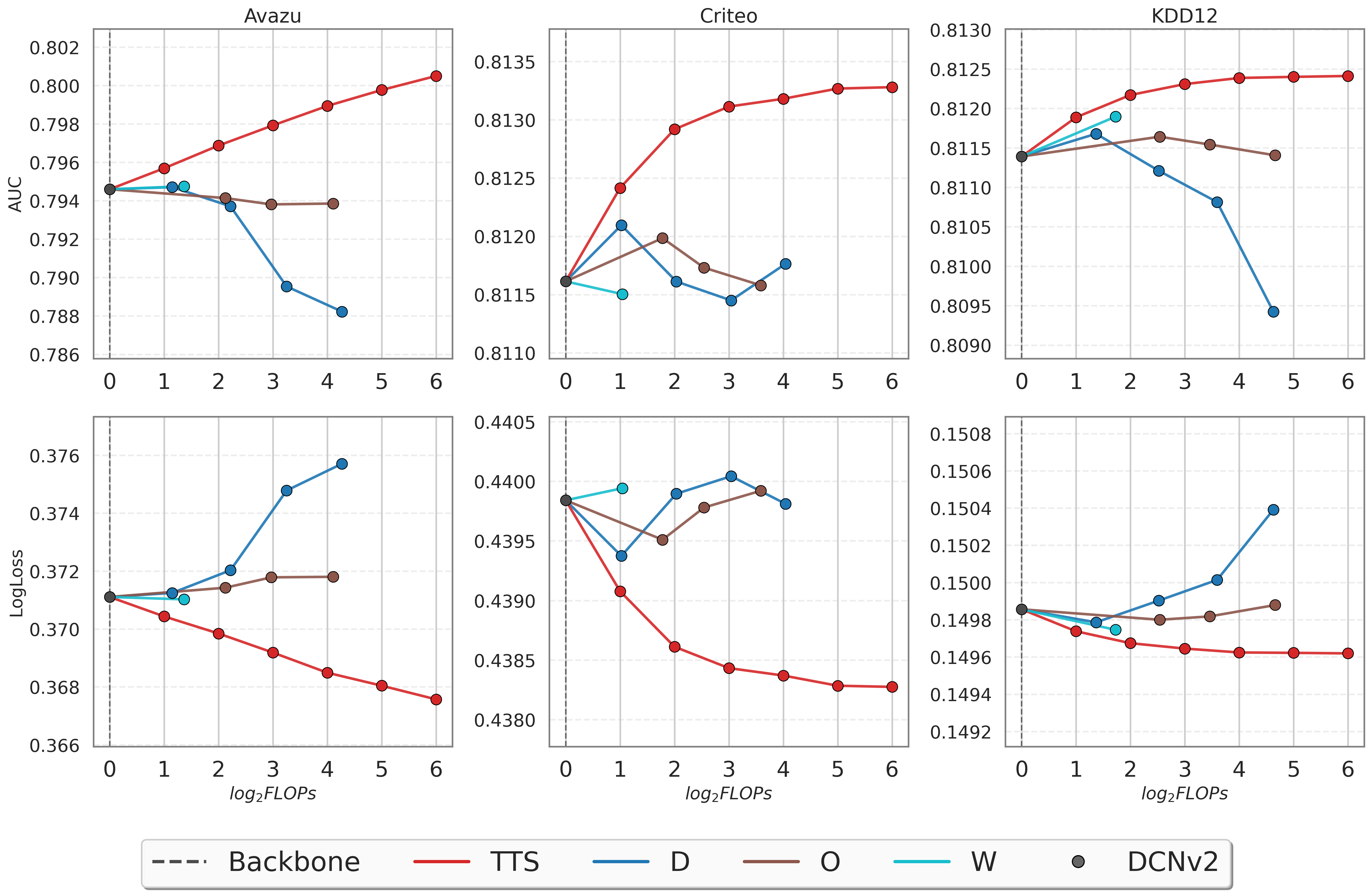}
        \caption{DCNv2}
        \label{fig:dcnv2}
    \end{subfigure}
    \hfill
    \begin{subfigure}[b]{0.48\textwidth}
        \centering
        \includegraphics[width=\linewidth]{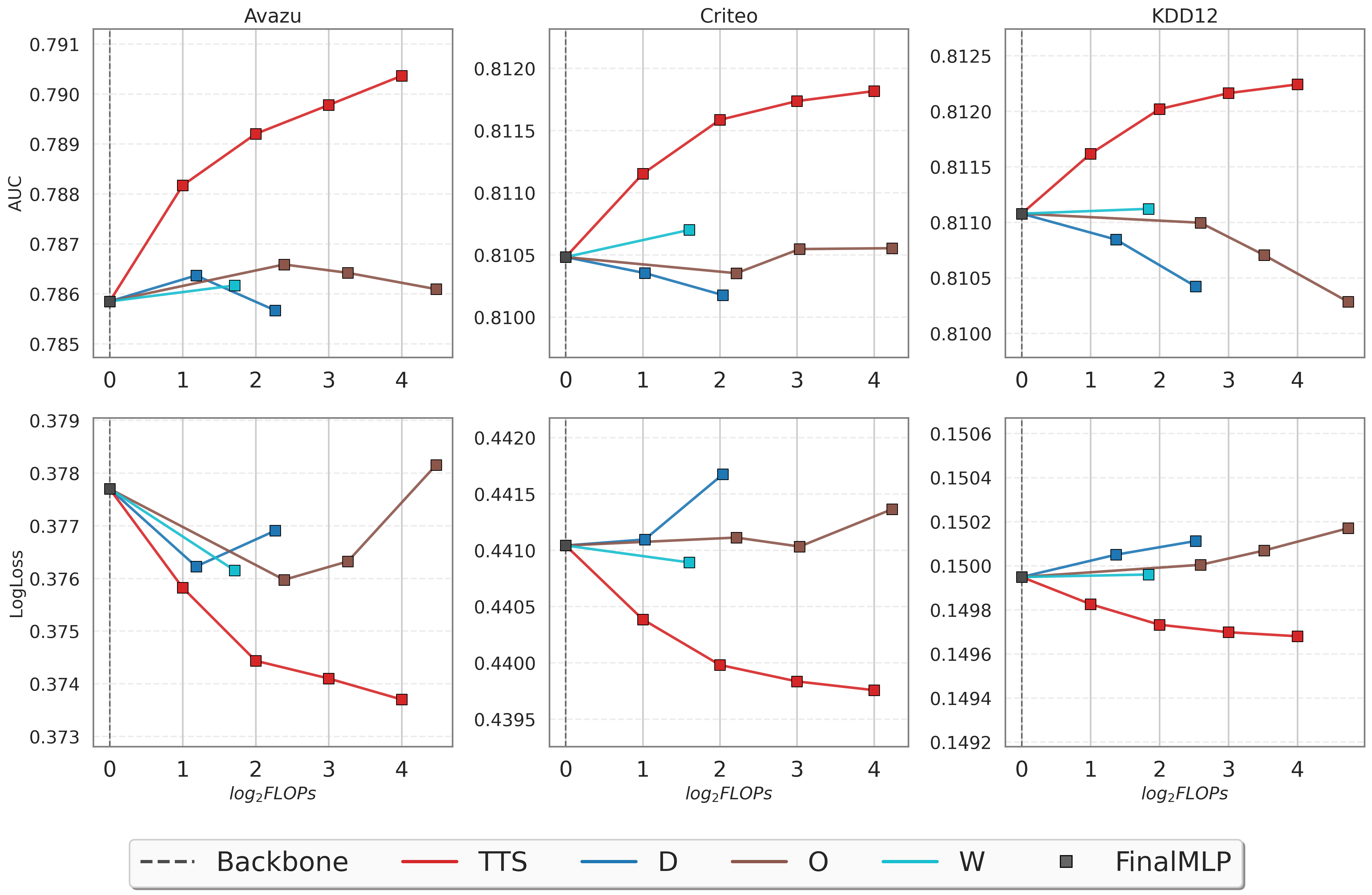}
        \caption{FinalMLP}
        \label{fig:finalmlp}
    \end{subfigure}
    \vspace{-10pt}
    \caption{Performance scaling curves across three datasets (Avazu, Criteo, and KDD12) as a function of multiplicative FLOPs scaling factor. Each subfigure shows a model's performance under different computational budgets. Scalable architectures (Wukong and RankMixer) exhibit consistent performance improvements following scaling laws, while others (DCNv2 and FinalMLP) demonstrate diminishing or negative returns at higher FLOPs, indicating limited parameter scalability.}
    \Description{Performance scaling curves for three datasets showing differential scalability across model architectures}
    \label{fig:performance-scaling}
    \vspace{-10pt}
\end{figure*}

\subsubsection{Efficiency}
In this section, we demonstrate that test-time scaling via prediction merging is, in fact, more efficient than depth scaling. Such an observation has been witnessed previously in the language model domain~\cite{optimal-compute}. Here we visualize both scaling solutions in the format of a performance-efficiency curve, shown in Figure~\ref{fig:performance-scaling}. Such an evaluation is conducted on both depth scaling models, namely Wukong and Rankmixer, and classic DL models, with DCNv2 and FinalMLP as representatives. For the parameter scaling, we illustrate various scaling dimensions when scaling up the single model parameters in Appendix~\ref{appendix:experiment_scaling_dim}. We can make the following observations: 
First, test-time scaling generally outperforms parameter scaling across all datasets and backbones, showcasing the superiority of our solution. 
Second, the performance of parameter scaling diverse across different datasets. For instance, Rankmixer demonstrates good parameter scalability on the Avazu and KDD12 datasets. Yet on the Criteo dataset, the scalability is not good. 
Finally, depth scaling models require extensive exploration on the scaling dimensions shown in Appendix~\ref{appendix:experiment_scaling_dim}. Such an effort, although not widely discussed in academia, is non-trivial when faced with large-scale production datasets. In contrast, test-time scaling does not rely on additional parameter tuning. 

\subsubsection{Stability}
We further compare the {stability} against depth scaling in Table~\ref{tab:stability}. Specifically, we train the depth-scaled models with 5 random seeds and measure the standard deviation. For the test-time scaling settings, we also calculate the standard deviation and repetitively sample 16$\times$ models from the model set trained on 64$\times$ random seeds for 5 times. We can observe that compared with depth scaling, test-time scaling yields more stable predictions in terms of both AUC and Logloss. This correlates with the advantage of test-time scaling in terms of stability.

\begin{table}[!htbp]
\centering
\caption{Standard Derivation over five random trials on Depth Scaling and Test-time Scaling.}
\vspace{-10pt}
\label{tab:stability}
\resizebox{0.48\textwidth}{!}{
\begin{tabular}{l|cc|cc}
\hline
    \multirow{2}{*}{Setup} & \multicolumn{2}{c|}{Depth Scaling} & \multicolumn{2}{c}{Test-time Scaling} \\
\cline{2-5}
    & $\sigma_{\text{AUC}}$ & $\sigma_{\text{Logloss}}$ & $\sigma_{\text{AUC}}$ & $\sigma_{\text{Logloss}}$ \\
\hline
    Avazu-Wukong (16x) &\num{2.99 E -4} &\num{1.7 E -4} & \num{7.05 E -5} & \num{3.87 E -5} \\
    Avazu-Rankmixer (16x) &\num{1.04 E -4} & \num{6.8 E -5} & \num{2.35 E -5} & \num{1.2 E -5} \\
    Criteo-Wukong (16x) &\num{6 E -5} & \num{1.15 E -4} & \num{2.11 E -5} & \num{2.49 E -5} \\
    Criteo-Rankmixer (16x) & \num{5.5 E -5} & \num{5.5 E -5} & \num{1.15 E -4} & \num{1.12 E -4} \\
    KDD12-Wukong (16x) & \num{1.97 E -4} & \num{4.6 E -5} & \num{3.41 E -5} & \num{9.5 E -6} \\
    KDD12-Rankmixer (16x) & \num{1.28 E -4} & \num{2.5 E -5} & \num{2.16 E -5} & \num{5.94 E -6} \\
\hline
\end{tabular}
}
\vspace{-10pt}
\end{table}

\subsection{RQ3: Comparison against Width Scaling}

In this section, we aim to investigate the effectiveness of prediction merging over width scaling. In previous width scaling methods~\cite{MultiEmbed,D-MoE}, the information fusion process typically occurs after the interaction stage. A shared predictor is adopted to make the final prediction based on the fused interactions. In contrast, our prediction merging tends to adopt late fusion and directly averages the predictions in the logits space. We compared MultiEmbed (ME)~\cite{MultiEmbed}, which can be viewed as a width scaling method without regularizations, and D-MoE~\cite{D-MoE}, which adopts expert-designed regularization, as baselines in this section. The further illustration of these baselines is described in Section~\ref{sec:method_param} and Figure~\ref{fig:method}. The results of applying depth scaling techniques to Wukong, Rankmixer, and their combinations are shown in Table~\ref{tab:moe}. 

\begin{table}[!htbp]
\centering
\caption{Comparison against Width Scaling}
\vspace{-10pt}
\label{tab:moe}
\resizebox{0.48\textwidth}{!}{
\begin{tabular}{c|c|cc|cc|cc}
\hline
    & \multirow{2}{*}{Model} & \multicolumn{2}{c|}{ME} & \multicolumn{2}{c|}{D-MoE} & \multicolumn{2}{c}{PM} \\
\cline{3-8}
    & & AUC$\uparrow$ & Logloss$\downarrow$ & AUC$\uparrow$ & Logloss$\downarrow$ & AUC$\uparrow$ & Logloss$\downarrow$ \\
\hline
\multirow{3}{*}{\rotatebox{90}{Avazu}}
    & W*2 & 0.7969 & \underline{0.3697} & \underline{0.7972} & 0.3709 & \textbf{0.7973} & \textbf{0.3693}$^*$ \\  
    & R*2 & 0.7949 & \underline{0.3708} & \underline{0.7954} & 0.3710 & \textbf{0.7964}$^*$ & \textbf{0.3699}$^*$ \\
    & R+W & 0.7953 & \underline{0.3707} & \underline{0.7976} & 0.3722 & \textbf{0.7987}$^*$ & \textbf{0.3684}$^*$ \\
\hline
\multirow{3}{*}{\rotatebox{90}{Criteo}}
    & W*2 & 0.8106 & 0.4410 & \underline{0.8111} & \underline{0.4408} & \textbf{0.8121}$^*$ & \textbf{0.4395}$^*$ \\
    & R*2 & \underline{0.8115} & 0.4406 & 0.8110 & \underline{0.4406} & \textbf{0.8124}$^*$ & \textbf{0.4392}$^*$ \\
    & R+W & \underline{0.8113} & \underline{0.4405} & 0.8110 & 0.4410 & \textbf{0.8129}$^*$ & \textbf{0.4387}$^*$ \\
\hline
\multirow{3}{*}{\rotatebox{90}{KDD12}}
    & W*2 & 0.8125 & \underline{0.1497} & \textbf{0.8130} & \underline{0.1497} & 0.8128 & \textbf{0.1496} \\
    & R*2 & 0.8126 & 0.1497 & \underline{0.8128} & \underline{0.1496} & \textbf{0.8133} & \textbf{0.1494} \\
    & R+W & 0.8119 & 0.1499 & \underline{0.8132} & \underline{0.1496} & \textbf{0.8133} & \textbf{0.1495} \\
\hline
\end{tabular}}
\begin{tablenotes}
    \footnotesize \item[1] \noindent Here, R and W are abbreviations of Rankmixer and Wukong, respectively. The best and second-best performed structure is highlighted in \textbf{bold} and \underline{underline} font, respectively. $^*$ denotes significant improvement between the best and second-best performed baselines, measured by a two-tailed t-test with $p$-value smaller than 0.05.
    \vspace{-10pt}
\end{tablenotes}
\end{table}

We can easily make the following observations:
First, we observe that Prediction Merging (PM) outperforms D-MoE and ME in the majority of the cases, with significant improvements consistently observed on Avazu and Criteo datasets.
Second, it can be observed that the ME and D-MoE methods do not yield stable improvements against a single model when compared to the data in Table~\ref{table:main_homo}. Specifically, on the Criteo dataset, increasing the number of interaction modules for both Wukong and Rankmixer would negatively affect the final performance.
Finally, when comparing D-MoE and ME, we can observe that the largest performance increase due to regularization happens when heterogeneous architectures are ensembled together, namely Wukong + Rankmixer. Such an observation is aligned with the conclusions in the original paper~\cite{D-MoE}.

\subsection{RQ4: Orthogonality Experiment}
\label{sec:experiment_orthogonal}
In this section, we aim to investigate whether the parameter scaling is orthogonal to test-time scaling. Specifically, we scale up both Wukong and Rankmixer, two models designed to be scalable, via the best-performing dimension. We specifically scale up the parameters by up to 16$\times$. For each scaled parameter model, we further scale via homogeneous models by 16$\times$. Note that we exclude the Criteo dataset, as it does not show a significant difference when scaling up parameters based on the result shown in Figure~\ref{fig:wukong} and~\ref{fig:rankmixer}. Based on Table~\ref{tab:ortho_scaling}, we can make the following observations. 

\begin{table*}[htbp]
\centering
\caption{Performance comparison of Wukong and Rankmixer with different scaling factors on Avazu and KDD12 datasets. The values in parentheses indicate the improvement over the corresponding Base model.}
\vspace{-10pt}
\label{tab:ortho_scaling}
\resizebox{0.98\textwidth}{!}{%
\begin{tabular}{c|c|c|cc|cc|cc|cc|cc}
    \hline
    \multirow{2}{*}{Dataset} & \multirow{2}{*}{Model} & \multirow{2}{*}{Scaling Type} & \multicolumn{2}{c|}{Param 1x} & \multicolumn{2}{c|}{Param 2x} & \multicolumn{2}{c|}{Param 4x} & \multicolumn{2}{c|}{Param 8x} & \multicolumn{2}{c}{Param 16x} \\
    \cline{4-13}
    & & & AUC$\uparrow$  & LogLoss$\downarrow$ & AUC$\uparrow$  & LogLoss$\downarrow$ & AUC$\uparrow$  & LogLoss$\downarrow$ & AUC$\uparrow$  & LogLoss$\downarrow$ & AUC$\uparrow$  & LogLoss$\downarrow$ \\
    \hline
    \multirow{4}{*}{\rotatebox{90}{Avazu}} 
        & \multirow{2}{*}{Wukong} & Base & 0.7954 & 0.3704 & 0.7972  & 0.3693 & 0.7979  & 0.3689  & 0.7983  & 0.3687  & 0.7984  & 0.3686  \\
        & & Test-time 16x & 0.7964 \tiny{(+10)} & 0.3698 \tiny{(-6)} & 0.7981 \tiny{(+9)} & 0.3688 \tiny{(-5)} & 0.7989 \tiny{(+10)} & 0.3683 \tiny{(-6)} & 0.7991 \tiny{(+8)} & 0.3681 \tiny{(-6)} & 0.7994 \tiny{(+10)} & 0.3680 \tiny{(-6)} \\
        \cline{2-13}
        & \multirow{2}{*}{Rankmixer} & Base & 0.7952 & 0.3706 & 0.7958  & 0.3702  & 0.7962  & 0.3700  & 0.7963  & 0.3699 & 0.7964  & 0.3698  \\
        & & Test-time 16x & 0.7964 \tiny{(+12)} & 0.3699 \tiny{(-7)} & 0.7968 \tiny{(+10)} & 0.3696 \tiny{(-6)} & 0.7970 \tiny{(+8)} & 0.3695 \tiny{(-5)} & 0.7971 \tiny{(+8)} & 0.3694 \tiny{(-5)} & 0.7971 \tiny{(+7)} & 0.3694 \tiny{(-14)} \\
    \hline
    \multirow{4}{*}{\rotatebox{90}{KDD12}} 
        & \multirow{2}{*}{Wukong} & Base & 0.8117 & 0.1498 & 0.8128  & 0.1496  & 0.8133  & 0.1495  & 0.8135  & 0.1494  & 0.8137  & 0.1494  \\
        & & Test-time 16x & 0.8118 \tiny{(+1)} & 0.1498 \tiny{(+0)} & 0.8130 \tiny{(+2)} & 0.1495 \tiny{(-1)} & 0.8135 \tiny{(+2)} & 0.1494 \tiny{(-1)} & 0.8138 \tiny{(+3)} & 0.1494 \tiny{(+0)} & 0.8139 \tiny{(+2)} & 0.1493 \tiny{(-1)} \\
        \cline{2-13}
        & \multirow{2}{*}{Rankmixer} & Base & 0.8124 & 0.1496 & 0.8131  & 0.1495  & 0.8134  & 0.1494  & 0.8135  & 0.1494  & 0.8136  & 0.1493  \\
        & & Test-time 16x & 0.8140 \tiny{(+16)} & 0.1493 \tiny{(-3)} & 0.8145 \tiny{(+14)} & 0.1492 \tiny{(-3)} & 0.8147 \tiny{(+13)} & 0.1491 \tiny{(-3)} & 0.8149 \tiny{(+14)} & 0.1491 \tiny{(-3)} & 0.8149 \tiny{(+13)} & 0.1491 \tiny{(-2)} \\
    \hline
\end{tabular}%
}
\vspace{-10pt}
\end{table*}

First, the parameter scaling shows consistent improvement. However, as the parameter increases, performance plateaus are commonly witnessed. Such an observation highlights the performance limitation of parameter scaling when the amount of training data is fixed. 
Second, compared with the base model, test-time scaling shows consistent improvement in both AUC and Logloss with scaling up. We can observe that the performance gap between the test-time scaled solution and the base model remains relatively stable when the parameters of the base model change. Such an observation highlights the orthogonality between test-time scaling and parameter scaling on large-scale recommendation models. A similar phenomenon is also observed on the LMs~\cite{optimal-compute,LLMMonkey}.

\subsection{RQ5: Comparison against Model Merging} \label{sec:experiment_merging}
Model merging has been an effective trick in language model domains. Specifically speaking, it fuses multiple foundation models with individually finetuned on different tasks, namely expert models~\cite{modeledit,TTMM,merging_survey,ties}. After fusion, these merged models can yield comparable performance on individual tasks compared with the corresponding expert models. In the meantime, the number of models and inference cost can be greatly reduced~\cite{TTMM}.

\begin{table}[!htbp]
\renewcommand\arraystretch{1.15}
\centering
\caption{Model vs. Prediction Merging.}
\label{table:merging_small}
\vspace{-10pt}
\resizebox{0.48\textwidth}{!}{
\begin{tabular}{c|c|cc|cc|cc|cc}
    \hline
    & \multirow{2}{*}{Method} & \multicolumn{2}{c|}{DCNv2} & \multicolumn{2}{c|}{FinalMLP} & \multicolumn{2}{c|}{Wukong} & \multicolumn{2}{c}{RankMixer} \\
    \cline{3-10}                            
    & & AUC$\uparrow$ & LogLoss$\downarrow$ & AUC$\uparrow$ & LogLoss$\downarrow$ & AUC$\uparrow$ & LogLoss$\downarrow$ & AUC$\uparrow$ & LogLoss$\downarrow$ \\
    \hline
    \multirow{7}{*}{\rotatebox{90}{Avazu}} 
    & Backbone-Model & 0.7944        & 0.3712              & 0.7858        & 0.3777              & 0.7954        & 0.3705              & 0.7952        & 0.3706         \\
    \cline{2-10}
    & 2$\times$-Model & 0.7707        & 0.5302              & 0.7681        & 0.5446              & 0.7152        & 0.6586              & 0.7689        & 0.6195         \\
    & 2$\times$-Pred  & 0.7952        & 0.3707              & 0.7882        & 0.3758              & 0.7972        & 0.3694              & 0.7962        & 0.3699         \\
    \cline{2-10}
    & 4$\times$-Model & 0.7447        & 0.6303              & 0.6377        & 0.6105              & 0.6942        & 0.6790              & 0.7289        & 0.6725         \\
    & 4$\times$-Pred  & 0.7956        & 0.3704              & 0.7892        & 0.3744              & 0.7986        & 0.3685              & 0.7971        & 0.3793         \\
    \cline{2-10}
    & 16$\times$-Model  & 0.7343        & 0.6832              & 0.5304        & 0.6483              & 0.4941        & 0.6847              & 0.4924        & 0.6829         \\
    & 16$\times$-Pred   & {0.7959} & {0.3702}   & {0.7904} & {0.3737}   & {0.8001} & {0.3676}   & {0.7988} & {0.3684} \\
    \hline
    \multirow{7}{*}{\rotatebox{90}{Criteo}} 
    & Backbone-Model   & 0.8116        & 0.4399              & 0.8105        & 0.4410              & 0.8103        & 0.4411              & 0.8115        & 0.4400         \\
    \cline{2-10}
    & 2$\times$-Model  & 0.7704        & 0.5480              & 0.7796        & 0.6071              & 0.7571        & 0.6507              & 0.7741        & 0.6387         \\
    & 2$\times$-Pred   & 0.8124        & 0.4391              & 0.8112        & 0.4404              & 0.8121        & 0.4395              & 0.8130        & 0.4386         \\
    \cline{2-10}
    & 4$\times$-Model  & 0.7472        & 0.6240              & 0.7405        & 0.6571              & 0.7066        & 0.6831              & 0.7242        & 0.6645         \\
    & 4$\times$-Pred   & 0.8129        & 0.4387              & 0.8116        & 0.4400              & 0.8131        & 0.4387              & 0.8138        & 0.4379         \\
    \cline{2-10}
    & 16$\times$-Model & 0.7254        & 0.6788              & 0.5000        & 0.6754              & 0.6109        & 0.6830              & 0.6812        & 0.6682         \\
    & 16$\times$-Pred  & 0.8132 & 0.4384   & 0.8118 & 0.4398   & 0.8137 & 0.4381   & 0.8143 & 0.4374 \\
    \hline
\end{tabular}
}

\begin{tablenotes}
    \footnotesize 
    \item \textbf{Model} denotes model weight merging (averaging parameters from models), while \textbf{Pred} denotes prediction merging (averaging outputs). 
\end{tablenotes}
\vspace{-10pt}
\end{table}

In this section, we elaborate on how such model merging techniques can be combined with large-scale recommendation models, which, different from LMs, are trained independently from scratch. Here, we adopt one of the commonly used techniques, which directly averages the weights among different models~\cite{merging_survey}. We report the performance of both prediction merging and model merging in Table~\ref{table:merging_small}. Observations are listed as follows.

First, prediction merging consistently outperforms model merging across all setups. This highlights the superiority of prediction merging over model merging in large-scale recommendation tasks.
Second, unlike prediction merging, where performance increases as the number of models increases, model merging exhibits a consistent decrease in performance in nearly all setups when the number of models is increased. As the number of models increases to 16$\times$, the performance of model merging may decrease to around 0.5 in terms of AUC, which renders the final results indistinguishable from random guessing. Such a surprising result highlights that model merging techniques may not be easily adopted in large-scale recommendation tasks. We believe this is likely due to the fact that in the language model domain, the merged model tends to share a common ancestor, a pre-trained foundation model that was not tuned on specific tasks. Hence, these parameter distribution may not differs too much from each other. However, in our setups, different experts are trained independently with individually random initialization. Hence, directly merging them can be hard without tailored techniques.

\begin{figure}[!htbp]
\centering
\begin{subfigure}[b]{0.30\textwidth}
    \centering
    \includegraphics[width=\linewidth]{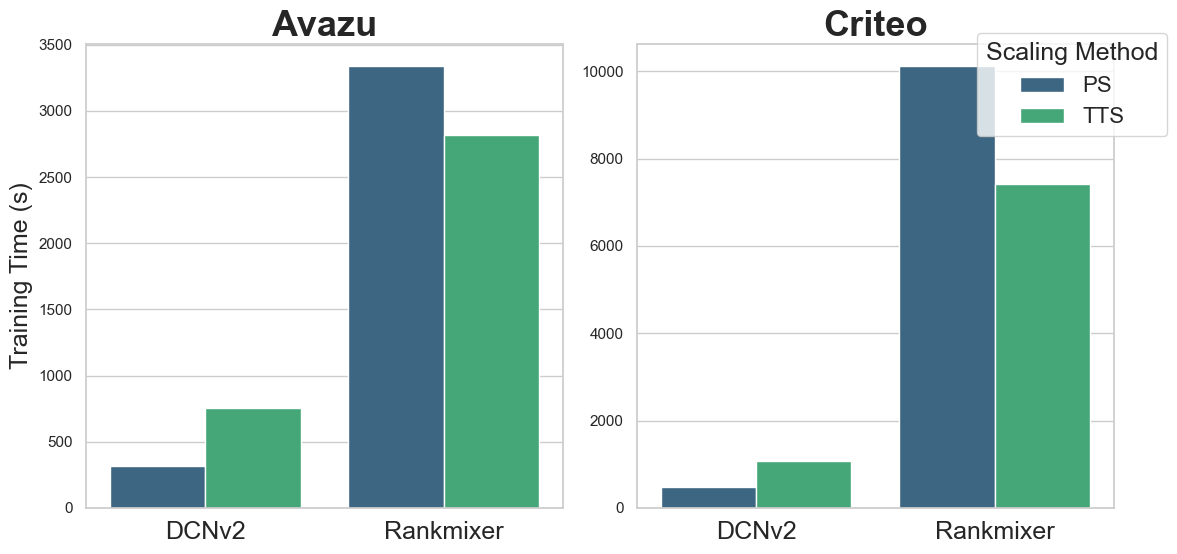}
    \caption{Training Time Per Epoch}
    \label{fig:efficiency-train}
\end{subfigure}
\hfill
\begin{subfigure}[b]{0.45\textwidth}
    \centering
    \includegraphics[width=\linewidth]{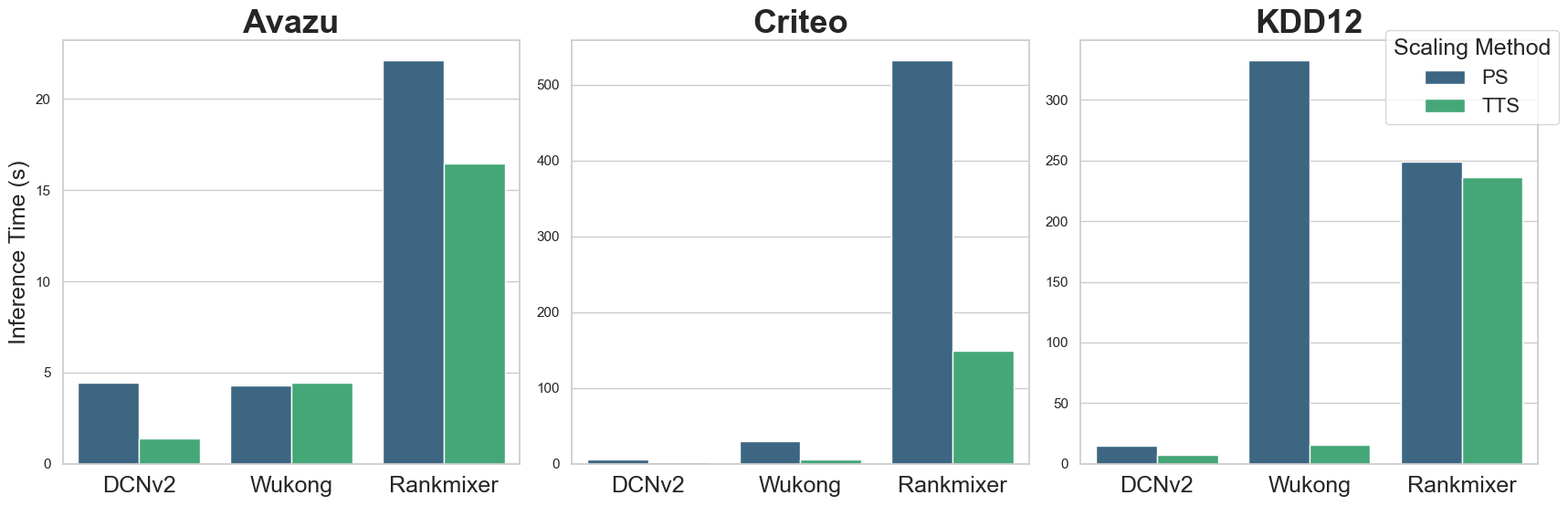}
    \caption{Inference Time on Validation}
    \label{fig:efficiency-inference}
\end{subfigure}
\vspace{-10pt}
\caption{Efficiency Study.}
\Description{Efficiency Study.}
\vspace{-15pt}
\label{fig:efficiency}
\end{figure}
\subsection{RQ6: Efficiency Study}

In this section, we aim to study the efficiency aspect of parameter scaling and test-time scaling. Specifically, we compare the following two groups: (a) parameter scaling with $8\times$, the best dimension in Figure~\ref{fig:performance-scaling} is selected, and (b) test-time scaling with $8\times$. Specifically, all models are run on an 8-core Intel(R) Xeon(R) Gold 6459C CPU, 40 GB of memory, and one A100 GPU with 48GB of GPU memory. For test-time scaling, all 8 models are trained and inferred on the same GPU with different processes. We plot the training time per epoch and inference time over the validation set in Figure~\ref{fig:efficiency-train} and~\ref{fig:efficiency-inference}, respectively. 

We can easily observe from Figure~\ref{fig:efficiency-inference} that test-time scaling can yield superior inference speed against parameter scaling in almost all datasets. Such an improvement is particularly evident in scalable models, namely Rankmixer and Wukong, as the interaction module, which sequentially concatenates, consumes a significant amount of time during inference. Additionally, test-time scaling approaches may require less training time on a scalable model, such as Rankmixer, compared to parameter scaling. Both observations prove the efficiency of test-time scaling in both training and inference stages. 
However, test-time scaling is surpassed by parameter scaling when trained on DCNv2 structures. This is likely because DCNv2 requires less computation than scalable models, making the I/O bandwidth a bottleneck during training. With more processes running at the same time, the training speed for test-time scaling models naturally reduced.

\subsection{RQ7: Case Study}

\begin{figure}[!htbp]
    \centering
    \begin{subfigure}[h]{0.15\textwidth}
        \centering
        \includegraphics[width=\linewidth]{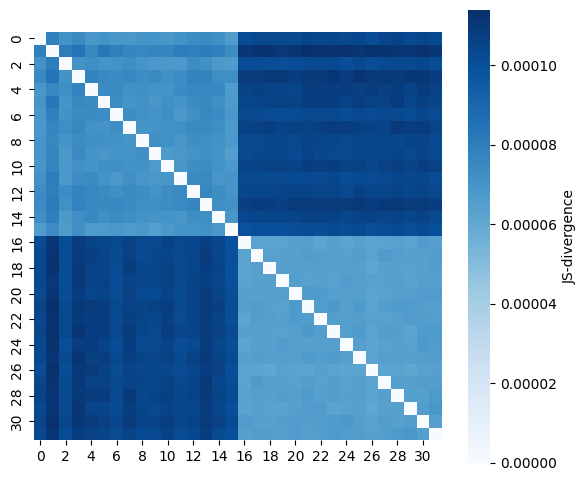}
        \caption{R+D on Avazu}
        \label{fig:case-avazu-r+d}
    \end{subfigure}
    \begin{subfigure}[h]{0.15\textwidth}
        \centering
        \includegraphics[width=\linewidth]{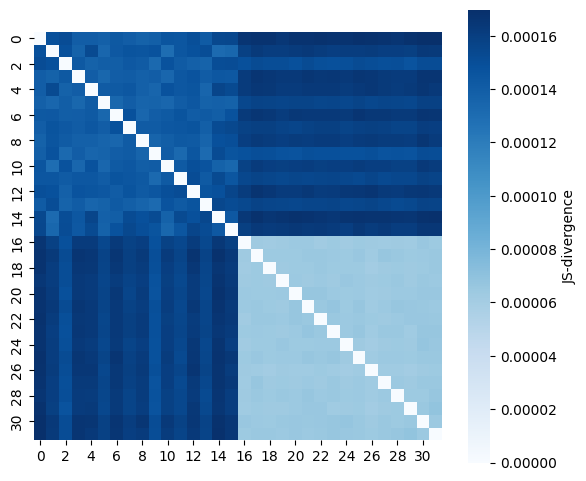}
        \caption{W+R on Avazu}
        \label{fig:case-avazu-w+r}
    \end{subfigure}
    \begin{subfigure}[h]{0.15\textwidth}
        \centering
        \includegraphics[width=\linewidth]{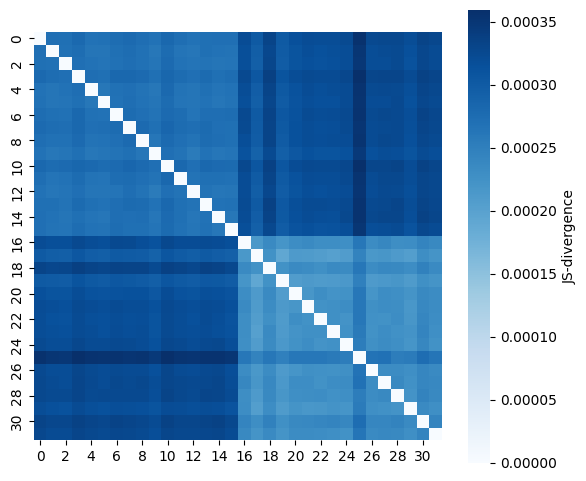}
        \caption{W+R on Criteo}
        \label{fig:case-criteo-w+r}
    \end{subfigure}

    \begin{subfigure}[b]{0.15\textwidth}
        \centering
        \includegraphics[width=\linewidth]{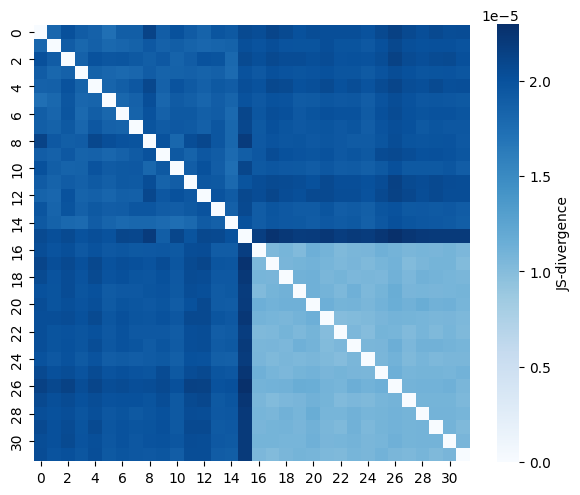}
        \caption{W+R on KDD12\vspace{-10pt}}
        \label{fig:case-kdd12-w+r}
    \end{subfigure}
    \begin{subfigure}[b]{0.22\textwidth}
        \centering
        \includegraphics[width=\linewidth]{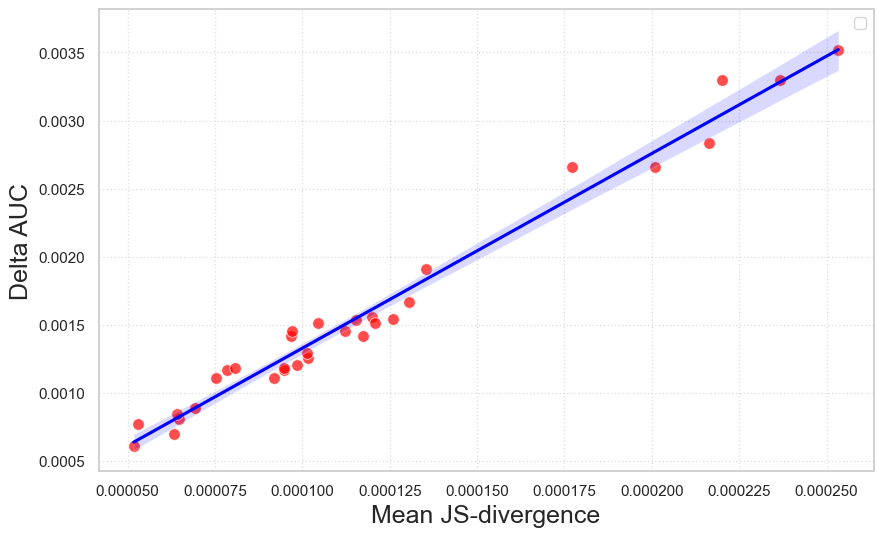}
        \caption{Criteo\vspace{-10pt}}
        \label{fig:case-criteo}
    \end{subfigure}
    \vspace{-5pt}
    \begin{minipage}[t]{0.45\columnwidth} 
    \centering
    \subcaption{JS-divergence of FinalMLP models on Avazu.} 
    \Description{JS-divergence of FinalMLP models on Avazu.}
    \includegraphics[width=\linewidth]{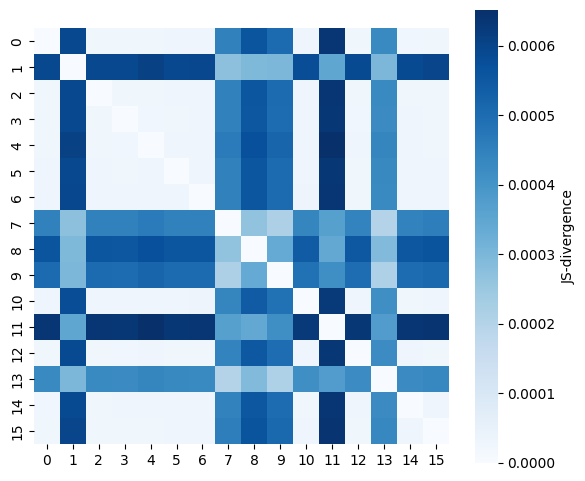} 
    \label{fig:case_finalmlp_figure}
    \end{minipage}%
    \hfill%
    \begin{minipage}[t]{0.45\columnwidth} 
    \centering
    \subcaption{Case Study over FinalMLP on Avazu Dataset. Here, Top-$k$ refers to selecting the top-$k$ models, measured by the average JS-divergence to other models.} 
    \label{tab:case_finalmlp_table}
    \resizebox{0.98\columnwidth}{!}{
    \begin{tabular}{l|cc}
    \hline
        Metric & ($\Delta$)AUC & ($\Delta$)Logloss \\
    \hline
        16$\times$ & 0.7904 & 0.3737 \\
        top-12 & +0.0006 & -0.0003 \\
        top-8 & +0.0012 & -0.0001 \\
        top-4 & -0.0001 & +0.0022 \\
    \hline
    \end{tabular}}
    \end{minipage}
    \vspace{-15pt}
    \caption{(a) -- (d): JS Divergence over Heterogeneous Architecture on Various Datasets. Here, D, W, and R are abbreviations for DCNv2, Wukong, and Rankmixer. For the name X+Y in the subcaption, X and Y refer to the first 16 indexes and last 16 indexes in the subfigure, where models X and Y are independently trained over 16 seeds, respectively. (e): Correlation between JS-divergence and $\Delta$AUC on Criteo. }
    \Description{Case Study over Heterogeneous Architecture on Various Datasets. Here, D, W, and R are abbreviations for DCNv2, Wukong, and Rankmixer.}
    \label{fig:case}
    \vspace{-10pt}
\end{figure}

In this section, we aim to investigate the reasons why test-time scaling can be achieved via prediction merging. As we hypothesize in the introduction, we emphasize that one of the key components in successfully applying test-time scaling is generating meaningful and diverse outputs. In this section, we visualize the Jensen–Shannon(JS) divergence among all pair combinations among models, as shown in Figure~\ref{fig:case}. Notes that we only include the JS divergence among heterogeneous architectures, as they naturally include homogenous architectures within them. The JS divergence between individual model pairs $i$ and $j$ is calculated as follows:
\begin{equation}
    \text{JS}(i,j) = \frac{1}{|\{\mathcal{X}_{test}, \mathcal{Y}_{test}\}|} \sum_{(x,y) \in \{\mathcal{X}_{test}, \mathcal{Y}_{test}\}} \text{JS}(\hat{y}^{(i)}, \hat{y}^{(j)}).
\end{equation}
The above value corresponds to the value in each figure with row $i$, column $j$ (and row $j$, column $j$ due to the symmetric nature of JS-divergence).
Based on the results in Figure~\ref{fig:case}, we can make the following conclusions.

Firstly, we can easily observe that diversity exists between fields in all cases. Combined with the results in Table~\ref{table:main_hetero}, we can observe that prediction merging generates non-trivial improvements against base models. 
Secondly, the diversity between heterogeneous architectures is much larger than that between homogenous architectures. This can be observed in all figures that the bottom-left and top-right corner tends to be darker than the top-left and bottom-right parts, corresponding to JS-divergence among heterogeneous architectures and among homogeneous architectures, respectively. Such an observation is also aligned with previous work~\cite{D-MoE,CETNet} and Table~\ref{table:main_hetero} that the highest improvement is commonly witnessed when combining strong baselines together.
Finally, it can be observed that Wukong + Rankmixer generates a higher degree of diversity than RankMixer + DCNv2 on the Avazu dataset. This partially explains the superior improvement when scaling up Wukong + Rankmixer on the Avazu datasets.

To further validate our hypothesis that diversity is the reason, we conducted two studies. First, we investigate the FinalMLP model over the Avazu dataset. Here we plot a similar JS-divergence heatmap in Figure~\ref{fig:case_finalmlp_figure}. It can be observed that certain models may generate diverse outputs when compared to other models. Therefore, instead of merging predictions from all 16 models, we select the Top-$k$ diverse models based on the following diversity score and only fuse the predictions generated from these models:
\begin{equation}
    \text{DS}(i) = \sum_{j \neq i} \text{JS}(i,j),
\end{equation}
where the $\text{DS}_i$ refers to the diverse score of model $i$ with, calculated by the JS-divergence of model $i$ against all other models. We report the changes in AUC and Logloss in Table~\ref{tab:case_finalmlp_table}, denoted as $\Delta$AUC and $\Delta$Logloss. We can observe that when $k=8$ or $12$, the selected fusion surprisingly improves performance compared to utilizing all predictions. Such an observation highlights the importance of diversity in prediction when fusing them. 
Second, we plot the correlation between mean JS-divergence over all scalings and $\Delta$AUC on the Criteo dataset, shown in Figure~\ref{fig:case-criteo}. A clear correlation exists between these two variables. This indicates that the improvement of AUC is correlated with the divergence in predictions.  


\section{Related Work}
\label{sec:related_work}
\subsection{Scaling Laws in Recommendation}
Since the exponential growth of LM driven by the scaling law~\cite{ScalingLaw}, scaling up recommendation models has become a trend in the field~\cite{SL4Rec,SL4AdsRetrieve}. In the recommendation model, these techniques can be categorized into two classes: sequential-based~\cite{HSTU,MTGR,OneRec,LUM,LSRM,ARGUS,SUAN,CLUE} and sequential-free~\cite{Wukong,RankMixer}. The sequence-based solutions tend to scale up the length of the sequence~\cite{HSTU,MTGR,OneRec} or focus on the sequence data~\cite{LUM,LSRM,ARGUS,SUAN}, while the sequence-free solutions are more focused on the general classification formulation and tend to increase the parameters of the existing model~\cite{Wukong,RankMixer,MultiEmbed,D-MoE,CETNet}. The sequence-free solutions can be further categorized into depth scaling~\cite{Wukong,RankMixer} and width scaling~\cite{MultiEmbed,CETNet,D-MoE}. Note that width scaling shares similarity with existing solutions that utilize various interaction components in recommendation~\cite{DeepFM,Wide_Deep,DCN,DCNv2,MMoE}. The major difference lies in the corresponding embedding table, which is used to avoid collisions and achieve better expressiveness~\cite{MultiEmbed}.  
Our paper is more correlated with the sequence-free solutions, as we do not hold any hypothesis on the training corpus. However, our proposed solutions are orthogonal to all existing parameter scaling solutions, as shown empirically in Section~\ref{sec:experiment_orthogonal}.



\subsection{Test-time Scaling}

Test-time scaling has consistently become a working solution in trading the inference speed for higher performance~\cite{TheBitterLesson}. Recently, test-time scaling has become a popular trend in the language model domain~\cite{s1,LLMMonkey,optimal-compute}. These solutions can generally be decomposed into two dimensions~\cite{optimal-compute,TTS-survey}. One is to scale up the length of the output token via solutions such as budget forcing~\cite{s1}. The other approach is to diversify the output in parallel and generate the final solution via techniques such as self-consistency~\cite{self-consistency} or best-of-N~\cite{LLMMonkey}. Our paper relates to the parallel category, as both of them aim to generate diverse outputs during inference.

\section{Conclusion}
\label{sec:conclusion}

In this paper, we systematically investigate the effectiveness of test-time scaling applied in large-scale recommendation systems. We identify the key challenge of applying test-time scaling in recommendation: generating diverse yet meaningful outputs. Hence, we propose two solutions for generating diverse outputs: heterogeneity among different recommendation models and randomness on homogenous architectures. Prediction merging is further proposed as a technique for merging outputs. Through evaluation on three datasets and eight backbones, we demonstrate the scalability of test-time scaling in large-scale recommendation systems. We further demonstrate that test-time scaling can be more efficient and stable than parameter scaling. Ablation further proves that the test-time scaling via prediction merging is orthogonal to parameter scaling. 


\normalem
\bibliographystyle{ACM-Reference-Format}
\bibliography{main}

@inproceedings{FM,
  author    = {Steffen Rendle},
  title     = {Factorization Machines},
  booktitle = {{ICDM} 2010, The 10th {IEEE} International Conference on Data Mining},
  pages     = {995--1000},
  publisher = {{IEEE} Computer Society},
  year      = {2010},
  address   = {Sydney, Australia},
  doi       = {10.1109/ICDM.2010.127}
}

@inproceedings{Wide_Deep,
  author    = {Heng{-}Tze Cheng and
               Levent Koc and
               Jeremiah Harmsen and
               Tal Shaked and
               Tushar Chandra and
               Hrishi Aradhye and
               Glen Anderson and
               Greg Corrado and
               Wei Chai and
               Mustafa Ispir and
               Rohan Anil and
               Zakaria Haque and
               Lichan Hong and
               Vihan Jain and
               Xiaobing Liu and
               Hemal Shah},
  title     = {Wide {\&} Deep Learning for Recommender Systems},
  booktitle = {Proceedings of the 1st Workshop on Deep Learning for Recommender Systems, DLRS@RecSys 2016},
  pages     = {7--10},
  publisher = {{ACM}},
  year      = {2016},
  address   = {Boston, MA, USA},
  doi       = {10.1145/2988450.2988454}
}

@inproceedings{FNN,
  author    = {Weinan Zhang and
               Tianming Du and
               Jun Wang},
  title     = {Deep Learning over Multi-field Categorical Data - - {A} Case Study on User Response Prediction},
  booktitle = {Advances in Information Retrieval - 38th European Conference on {IR} Research, {ECIR} 2016},
  volume    = {9626},
  pages     = {45--57},
  publisher = {Springer},
  year      = {2016},
  address   = {Padua, Italy},
  doi       = {10.1007/978-3-319-30671-14}
}

@inproceedings{DeepFM,
  author    = {Huifeng Guo and
               Ruiming Tang and
               Yunming Ye and
               Zhenguo Li and
               Xiuqiang He},
  title     = {DeepFM: {A} Factorization-Machine based Neural Network for {CTR} Prediction},
  booktitle = {26th International Joint Conference on Artificial Intelligence, {IJCAI} 2017},
  pages     = {1725--1731},
  publisher = {ijcai.org},
  address   = {Melbourne, Australia},
  year      = {2017},
}

@inproceedings{IPNN,  
  author    = {Qu, Yanru and 
               Cai, Han and 
               Ren, Kan and 
               Zhang, Weinan and 
               Yu, Yong and 
               Wen, Ying and 
               Wang, Jun},  
  booktitle = {2016 IEEE 16th International Conference on Data Mining (ICDM)},  
  title     = {Product-Based Neural Networks for User Response Prediction},   
  address   = {Barcelona, Spain},
  publisher = {IEEE},
  year      = {2016},  
  pages     = {1149-1154},  
  doi       = {10.1109/ICDM.2016.0151}
}

@article{PIN,
  author    = {Yanru Qu and
               Bohui Fang and
               Weinan Zhang and
               Ruiming Tang and
               Minzhe Niu and
               Huifeng Guo and
               Yong Yu and
               Xiuqiang He},
  title     = {Product-Based Neural Networks for User Response Prediction over Multi-Field Categorical Data},
  journal   = {{ACM} Trans. Inf. Syst.},
  volume    = {37},
  number    = {1},
  pages     = {5:1--5:35},
  year      = {2019}
}

@inproceedings{xDeepFM,
  author    = {Jianxun Lian and
               Xiaohuan Zhou and
               Fuzheng Zhang and
               Zhongxia Chen and
               Xing Xie and
               Guangzhong Sun},
  title     = {xDeepFM: Combining Explicit and Implicit Feature Interactions for Recommender Systems},
  booktitle = {Proceedings of the 24th {ACM} {SIGKDD} International Conference on Knowledge Discovery {\&} Data Mining, {KDD} 2018},
  pages     = {1754--1763},
  address   = {London, UK},
  publisher = {{ACM}},
  year      = {2018},
  doi       = {10.1145/3219819.3220023},
}

@inproceedings{DCN,
  author       = {Ruoxi Wang and
                  Bin Fu and
                  Gang Fu and
                  Mingliang Wang},
  title        = {Deep {\&} Cross Network for Ad Click Predictions},
  booktitle    = {Proceedings of the ADKDD'17},
  pages        = {12:1--12:7},
  publisher    = {{ACM}},
  address      = {Halifax, NS, Canada},
  year         = {2017},
  url          = {https://doi.org/10.1145/3124749.3124754},
  doi          = {10.1145/3124749.3124754},
  bibsource    = {dblp computer science bibliography, https://dblp.org}
}

@inproceedings{DCNv2,
  author       = {Ruoxi Wang and
                  Rakesh Shivanna and
                  Derek Zhiyuan Cheng and
                  Sagar Jain and
                  Dong Lin and
                  Lichan Hong and
                  Ed H. Chi},
  title        = {{DCN} {V2:} Improved Deep {\&} Cross Network and Practical Lessons for Web-scale Learning to Rank Systems},
  booktitle    = {{WWW} '21: The Web Conference 2021},
  pages        = {1785--1797},
  address      = {Virtual Event / Ljubljana, Slovenia},
  publisher    = {{ACM} / {IW3C2}},
  year         = {2021},
  doi          = {10.1145/3442381.3450078},
}

@inproceedings{FinalMLP,
  author       = {Kelong Mao and
                  Jieming Zhu and
                  Liangcai Su and
                  Guohao Cai and
                  Yuru Li and
                  Zhenhua Dong},
  title        = {FinalMLP: An Enhanced Two-Stream {MLP} Model for {CTR} Prediction},
  booktitle    = {Thirty-Seventh {AAAI} Conference on Artificial Intelligence, {AAAI} 2023},
  pages        = {4552--4560},
  address      = {Washington, DC, USA},
  publisher    = {{AAAI} Press},
  year         = {2023},
  doi          = {10.1609/AAAI.V37I4.25577},
}

@article{DLRM,
  author       = {Maxim Naumov and
                  Dheevatsa Mudigere and
                  Hao{-}Jun Michael Shi and
                  Jianyu Huang and
                  Narayanan Sundaraman and
                  Jongsoo Park and
                  Xiaodong Wang and
                  Udit Gupta and
                  Carole{-}Jean Wu and
                  Alisson G. Azzolini and
                  Dmytro Dzhulgakov and
                  Andrey Mallevich and
                  Ilia Cherniavskii and
                  Yinghai Lu and
                  Raghuraman Krishnamoorthi and
                  Ansha Yu and
                  Volodymyr Kondratenko and
                  Stephanie Pereira and
                  Xianjie Chen and
                  Wenlin Chen and
                  Vijay Rao and
                  Bill Jia and
                  Liang Xiong and
                  Misha Smelyanskiy},
  title        = {Deep Learning Recommendation Model for Personalization and Recommendation Systems},
  journal      = {CoRR},
  volume       = {abs/1906.00091},
  year         = {2019},
  address      = {},
  pages        = {},
  url          = {http://arxiv.org/abs/1906.00091},
}

@inproceedings{OptFS,
  author       = {Fuyuan Lyu and
                  Xing Tang and
                  Dugang Liu and
                  Liang Chen and
                  Xiuqiang He and
                  Xue Liu},
  title        = {Optimizing Feature Set for Click-Through Rate Prediction},
  booktitle    = {Proceedings of the {ACM} Web Conference 2023, {WWW} 2023},
  pages        = {3386--3395},
  publisher    = {{ACM}},
  year         = {2023},
  address      = {Austin, TX, USA},
  doi          = {10.1145/3543507.3583545},
}

@inproceedings{AutoField,
  author    = {Yejing Wang and
               Xiangyu Zhao and
               Tong Xu and
               Xian Wu},
  title     = {AutoField: Automating Feature Selection in Deep Recommender Systems},
  booktitle = {{WWW} '22: The {ACM} Web Conference 2022},
  pages     = {1977--1986},
  publisher = {{ACM}},
  year      = {2022},
  address   = {Virtual Event, Lyon, France},
  doi       = {10.1145/3485447.3512071}
}

@misc{Criteo-dataset,
    author = {Jean-Baptiste Tien, joycenv, Olivier Chapelle},
    title = {Display Advertising Challenge},
    publisher = {Kaggle},
    year = {2014},
    url = {https://kaggle.com/competitions/criteo-display-ad-challenge}
}

@misc{Avazu-dataset,
    author = {Steve Wang and Will Cukierski},
    title = {Click-Through Rate Prediction},
    year = {2014},
    howpublished = {\url{https://kaggle.com/competitions/avazu-ctr-prediction}},
    note = {Kaggle}
}

@misc{KDD12-dataset,
    author = {Aden and Yi Wang},
    title = {KDD Cup 2012, Track 2},
    year = {2012},
    howpublished = {\url{https://kaggle.com/competitions/kddcup2012-track2}},
    note = {Kaggle}
}

@article{D-MoE,
  author       = {Jiancheng Wang and
                  Mingjia Yin and
                  Junwei Pan and
                  Ximei Wang and
                  Hao Wang and
                  Enhong Chen},
  title        = {Enhancing {CTR} Prediction with De-correlated Expert Networks},
  journal      = {CoRR},
  volume       = {abs/2505.17925},
  year         = {2025},
  url          = {https://doi.org/10.48550/arXiv.2505.17925},
  doi          = {10.48550/ARXIV.2505.17925},
  eprinttype    = {arXiv},
  pages        = {},
  eprint       = {2505.17925},
}

@inproceedings{MultiEmbed,
  author       = {Xingzhuo Guo and
                  Junwei Pan and
                  Ximei Wang and
                  Baixu Chen and
                  Jie Jiang and
                  Mingsheng Long},
  title        = {On the Embedding Collapse when Scaling up Recommendation Models},
  booktitle    = {Forty-first International Conference on Machine Learning, {ICML} 2024},
  publisher    = {OpenReview.net},
  address      = {Vienna, Austria},
  year         = {2024},
  pages        = {},
  url          = {https://openreview.net/forum?id=aPVwOAr1aW},
}

@article{CETNet,
  author       = {Xiaolong Liu and
                  Zhichen Zeng and
                  Xiaoyi Liu and
                  Siyang Yuan and
                  Weinan Song and
                  Mengyue Hang and
                  Yiqun Liu and
                  Chaofei Yang and
                  Donghyun Kim and
                  Wen{-}Yen Chen and
                  Jiyan Yang and
                  Yiping Han and
                  Rong Jin and
                  Bo Long and
                  Hanghang Tong and
                  Philip S. Yu},
  title        = {A Collaborative Ensemble Framework for {CTR} Prediction},
  journal      = {CoRR},
  volume       = {abs/2411.13700},
  year         = {2024},
  doi          = {10.48550/ARXIV.2411.13700},
  eprinttype    = {arXiv},
  pages        = {},
  eprint       = {2411.13700},
}

@inproceedings{MMoE,
  author       = {Jiaqi Ma and
                  Zhe Zhao and
                  Xinyang Yi and
                  Jilin Chen and
                  Lichan Hong and
                  Ed H. Chi},
  title        = {Modeling Task Relationships in Multi-task Learning with Multi-gate Mixture-of-Experts},
  booktitle    = {Proceedings of the 24th {ACM} {SIGKDD} International Conference on Knowledge Discovery {\&} Data Mining, {KDD} 2018},
  pages        = {1930--1939},
  address      = {London, UK},
  publisher    = {{ACM}},
  year         = {2018},
  doi          = {10.1145/3219819.3220007},
}

@inproceedings{MoE,
  author       = {Noam Shazeer and
                  Azalia Mirhoseini and
                  Krzysztof Maziarz and
                  Andy Davis and
                  Quoc V. Le and
                  Geoffrey E. Hinton and
                  Jeff Dean},
  title        = {Outrageously Large Neural Networks: The Sparsely-Gated Mixture-of-Experts Layer},
  booktitle    = {5th International Conference on Learning Representations, {ICLR} 2017},
  address      = {Toulon, France},
  publisher    = {OpenReview.net},
  year         = {2017},
  pages        = {},
  url          = {https://openreview.net/forum?id=B1ckMDqlg},
  bibsource    = {dblp computer science bibliography, https://dblp.org}
}

@article{ScalingLaw,
  author       = {Jared Kaplan and
                  Sam McCandlish and
                  Tom Henighan and
                  Tom B. Brown and
                  Benjamin Chess and
                  Rewon Child and
                  Scott Gray and
                  Alec Radford and
                  Jeffrey Wu and
                  Dario Amodei},
  title        = {Scaling Laws for Neural Language Models},
  journal      = {CoRR},
  volume       = {abs/2001.08361},
  year         = {2020},
  pages        = {},
  url          = {https://arxiv.org/abs/2001.08361},
}

@article{TheBitterLesson,
  title={The bitter lesson},
  author={Sutton, Richard},
  journal={Incomplete Ideas (blog)},
  volume={13},
  number={1},
  pages={38},
  year={2019}
}

@article{TTS-survey,
  author       = {Qiyuan Zhang and
                  Fuyuan Lyu and
                  Zexu Sun and
                  Lei Wang and
                  Weixu Zhang and
                  Zhihan Guo and
                  Yufei Wang and
                  Irwin King and
                  Xue Liu and
                  Chen Ma},
  title        = {What, How, Where, and How Well? {A} Survey on Test-Time Scaling in Large Language Models},
  journal      = {CoRR},
  volume       = {abs/2503.24235},
  year         = {2025},
  url          = {https://doi.org/10.48550/arXiv.2503.24235},
  doi          = {10.48550/ARXIV.2503.24235},
  eprinttype    = {arXiv},
  pages        = {},
  eprint       = {2503.24235},
}

@article{s1,
  author       = {Niklas Muennighoff and
                  Zitong Yang and
                  Weijia Shi and
                  Xiang Lisa Li and
                  Li Fei{-}Fei and
                  Hannaneh Hajishirzi and
                  Luke Zettlemoyer and
                  Percy Liang and
                  Emmanuel J. Cand{\`{e}}s and
                  Tatsunori Hashimoto},
  title        = {s1: Simple test-time scaling},
  journal      = {CoRR},
  volume       = {abs/2501.19393},
  year         = {2025},
  pages        = {},
  url          = {https://doi.org/10.48550/arXiv.2501.19393},
  doi          = {10.48550/ARXIV.2501.19393},
}

@article{optimal-compute,
  author       = {Charlie Snell and
                  Jaehoon Lee and
                  Kelvin Xu and
                  Aviral Kumar},
  title        = {Scaling {LLM} Test-Time Compute Optimally can be More Effective than caling Model Parameters},
  journal      = {CoRR},
  volume       = {abs/2408.03314},
  year         = {2024},
  pages        = {},
  url          = {https://doi.org/10.48550/arXiv.2408.03314},
  doi          = {10.48550/ARXIV.2408.03314},
}

@article{LLMMonkey,
  author       = {Bradley C. A. Brown and
                  Jordan Juravsky and
                  Ryan Ehrlich and
                  Ronald Clark and
                  Quoc V. Le and
                  Christopher R{\'{e}} and
                  Azalia Mirhoseini},
  title        = {Large Language Monkeys: Scaling Inference Compute with Repeated Sampling},
  journal      = {CoRR},
  volume       = {abs/2407.21787},
  year         = {2024},
  pages        = {},
  doi          = {10.48550/ARXIV.2407.21787},
  eprinttype    = {arXiv},
  eprint       = {2407.21787},
}

@inproceedings{self-consistency,
  author       = {Xuezhi Wang and
                  Jason Wei and
                  Dale Schuurmans and
                  Quoc V. Le and
                  Ed H. Chi and
                  Sharan Narang and
                  Aakanksha Chowdhery and
                  Denny Zhou},
  title        = {Self-Consistency Improves Chain of Thought Reasoning in Language Models},
  booktitle    = {The Eleventh International Conference on Learning Representations, {ICLR} 2023},
  pages        = {},
  publisher    = {OpenReview.net},
  address      = {Kigali, Rwanda},
  year         = {2023},
  url          = {https://openreview.net/forum?id=1PL1NIMMrw},
}

@inproceedings{lth,
  author       = {Jonathan Frankle and
                  Michael Carbin},
  title        = {The Lottery Ticket Hypothesis: Finding Sparse, Trainable Neural Networks},
  booktitle    = {7th International Conference on Learning Representations, {ICLR} 2019},
  publisher    = {OpenReview.net},
  address      = {New Orleans, LA, USA},
  pages        = {},
  year         = {2019},
  url          = {https://openreview.net/forum?id=rJl-b3RcF7}
}

@inproceedings{Wukong,
  author       = {Buyun Zhang and
                  Liang Luo and
                  Yuxin Chen and
                  Jade Nie and
                  Xi Liu and
                  Shen Li and
                  Yanli Zhao and
                  Yuchen Hao and
                  Yantao Yao and
                  Ellie Dingqiao Wen and
                  Jongsoo Park and
                  Maxim Naumov and
                  Wenlin Chen},
  title        = {Wukong: Towards a Scaling Law for Large-Scale Recommendation},
  booktitle    = {Forty-first International Conference on Machine Learning, {ICML} 2024},
  publisher    = {OpenReview.net},
  address      = {Vienna, Austria},
  year         = {2024},
  url          = {https://openreview.net/forum?id=8iUgr2nuwo},
  pages        = {},
}

@article{RankMixer,
  author       = {Jie Zhu and
                  Zhifang Fan and
                  Xiaoxie Zhu and
                  Yuchen Jiang and
                  Hangyu Wang and
                  Xintian Han and
                  Haoran Ding and
                  Xinmin Wang and
                  Wenlin Zhao and
                  Zhen Gong and
                  Huizhi Yang and
                  Zheng Chai and
                  Zhe Chen and
                  Yuchao Zheng and
                  Qiwei Chen and
                  Feng Zhang and
                  Xun Zhou and
                  Peng Xu and
                  Xiao Yang and
                  Di Wu and
                  Zuotao Liu},
  title        = {RankMixer: Scaling Up Ranking Models in Industrial Recommenders},
  journal      = {CoRR},
  volume       = {abs/2507.15551},
  year         = {2025},
  doi          = {10.48550/ARXIV.2507.15551},
  eprinttype    = {arXiv},
  eprint       = {2507.15551},
  pages        = {},
}

@inproceedings{SUAN,
  author       = {Weijiang Lai and
                  Beihong Jin and
                  Jiongyan Zhang and
                  Yiyuan Zheng and
                  Jian Dong and
                  Jia Cheng and
                  Jun Lei and
                  Xingxing Wang},
  title        = {Exploring Scaling Laws of {CTR} Model for Online Performance Improvement},
  booktitle    = {Proceedings of the Nineteenth {ACM} Conference on Recommender Systems, RecSys 2025},
  pages        = {114--123},
  address      = {Prague, Czech Republic},
  publisher    = {{ACM}},
  year         = {2025},
  doi          = {10.1145/3705328.3748046},
}

@inproceedings{CLUE,
  author       = {Kyuyong Shin and
                  Hanock Kwak and
                  Su Young Kim and
                  Max Nihl{\'{e}}n Ramstr{\"{o}}m and
                  Jisu Jeong and
                  Jung{-}Woo Ha and
                  Kyung{-}Min Kim},
  title        = {Scaling Law for Recommendation Models: Towards General-Purpose User Representations},
  booktitle    = {Thirty-Seventh {AAAI} Conference on Artificial Intelligence, {AAAI} 2023},
  pages        = {4596--4604},
  publisher    = {{AAAI} Press},
  address      = {Washington, DC, USA},
  year         = {2023},
  doi          = {10.1609/AAAI.V37I4.25582},
}

@inproceedings{HSTU,
  author       = {Jiaqi Zhai and
                  Lucy Liao and
                  Xing Liu and
                  Yueming Wang and
                  Rui Li and
                  Xuan Cao and
                  Leon Gao and
                  Zhaojie Gong and
                  Fangda Gu and
                  Jiayuan He and
                  Yinghai Lu and
                  Yu Shi},
  title        = {Actions Speak Louder than Words: Trillion-Parameter Sequential Transducers for Generative Recommendations},
  booktitle    = {Forty-first International Conference on Machine Learning, {ICML} 2024},
  publisher    = {OpenReview.net},
  address      = {Vienna, Austria},
  year         = {2024},
  pages        = {},
  url          = {https://openreview.net/forum?id=xye7iNsgXn},
}

@article{MTGR,
  author       = {Ruidong Han and
                  Bin Yin and
                  Shangyu Chen and
                  He Jiang and
                  Fei Jiang and
                  Xiang Li and
                  Chi Ma and
                  Mincong Huang and
                  Xiaoguang Li and
                  Chunzhen Jing and
                  Yueming Han and
                  Menglei Zhou and
                  Lei Yu and
                  Chuan Liu and
                  Wei Lin},
  title        = {{MTGR:} Industrial-Scale Generative Recommendation Framework in Meituan},
  journal      = {CoRR},
  volume       = {abs/2505.18654},
  year         = {2025},
  doi          = {10.48550/ARXIV.2505.18654},
  eprinttype    = {arXiv},
  pages        = {},
  eprint       = {2505.18654},
}

@article{OneRec,
  author       = {Jiaxin Deng and
                  Shiyao Wang and
                  Kuo Cai and
                  Lejian Ren and
                  Qigen Hu and
                  Weifeng Ding and
                  Qiang Luo and
                  Guorui Zhou},
  title        = {OneRec: Unifying Retrieve and Rank with Generative Recommender and Iterative Preference Alignment},
  journal      = {CoRR},
  volume       = {abs/2502.18965},
  year         = {2025},
  doi          = {10.48550/ARXIV.2502.18965},
  eprinttype    = {arXiv},
  pages        = {},
  eprint       = {2502.18965},
}

@inproceedings{LSRM,
  author       = {Gaowei Zhang and
                  Yupeng Hou and
                  Hongyu Lu and
                  Yu Chen and
                  Wayne Xin Zhao and
                  Ji{-}Rong Wen},
  title        = {Scaling Law of Large Sequential Recommendation Models},
  booktitle    = {Proceedings of the 18th {ACM} Conference on Recommender Systems, RecSys 2024},
  pages        = {444--453},
  address      = {Bari, Italy},
  publisher    = {{ACM}},
  year         = {2024},
  doi          = {10.1145/3640457.3688129},
}

@article{SL4Rec,
  author       = {Newsha Ardalani and
                  Carole{-}Jean Wu and
                  Zeliang Chen and
                  Bhargav Bhushanam and
                  Adnan Aziz},
  title        = {Understanding Scaling Laws for Recommendation Models},
  journal      = {CoRR},
  volume       = {abs/2208.08489},
  year         = {2022},
  doi          = {10.48550/ARXIV.2208.08489},
  eprinttype    = {arXiv},
  pages        = {},
  eprint       = {2208.08489},
}

@article{SL4AdsRetrieve,
  author       = {Yunli Wang and
                  Zixuan Yang and
                  Zhen Zhang and
                  Zhiqiang Wang and
                  Jian Yang and
                  Shiyang Wen and
                  Peng Jiang and
                  Kun Gai},
  title        = {Scaling Laws for Online Advertisement Retrieval},
  journal      = {CoRR},
  volume       = {abs/2411.13322},
  year         = {2024},
  doi          = {10.48550/ARXIV.2411.13322},
  eprinttype    = {arXiv},
  pages        = {},
  eprint       = {2411.13322},
}

@article{LUM,
  author       = {Bencheng Yan and
                  Shilei Liu and
                  Zhiyuan Zeng and
                  Zihao Wang and
                  Yizhen Zhang and
                  Yujin Yuan and
                  Langming Liu and
                  Jiaqi Liu and
                  Di Wang and
                  Wenbo Su and
                  Pengjie Wang and
                  Jian Xu and
                  Bo Zheng},
  title        = {Unlocking Scaling Law in Industrial Recommendation Systems with a Three-step Paradigm based Large User Model},
  journal      = {CoRR},
  volume       = {abs/2502.08309},
  year         = {2025},
  doi          = {10.48550/ARXIV.2502.08309},
  eprinttype    = {arXiv},
  pages        = {},
  eprint       = {2502.08309},
}

@article{ARGUS,
  author       = {Kirill Khrylchenko and
                  Artem Matveev and
                  Sergey S. Makeev and
                  Vladimir Baikalov},
  title        = {Scaling Recommender Transformers to One Billion Parameters},
  journal      = {CoRR},
  volume       = {abs/2507.15994},
  year         = {2025},
  doi          = {10.48550/ARXIV.2507.15994},
  eprinttype    = {arXiv},
  pages        = {},
  eprint       = {2507.15994},
}

@inproceedings{ties,
  author       = {Prateek Yadav and
                  Derek Tam and
                  Leshem Choshen and
                  Colin A. Raffel and
                  Mohit Bansal},
  title        = {TIES-Merging: Resolving Interference When Merging Models},
  booktitle    = {Advances in Neural Information Processing Systems 36: Annual Conference
                  on Neural Information Processing Systems 2023, NeurIPS 2023},
  year         = {2023},
  pages        = {},
  publisher    = {Curran Associates},
  address      = {New Orleans, LA, USA},
  url          = {http://papers.nips.cc/paper\_files/paper/2023/hash/1644c9af28ab7916874f6fd6228a9bcf-Abstract-Conference.html},
}

@inproceedings{modeledit,
  author       = {Gabriel Ilharco and
                  Marco T{\'{u}}lio Ribeiro and
                  Mitchell Wortsman and
                  Ludwig Schmidt and
                  Hannaneh Hajishirzi and
                  Ali Farhadi},
  title        = {Editing models with task arithmetic},
  booktitle    = {The Eleventh International Conference on Learning Representations,
                  {ICLR} 2023},
  publisher    = {OpenReview.net},
  address      = {Kigali, Rwanda},
  pages        = {},
  year         = {2023},
  url          = {https://openreview.net/forum?id=6t0Kwf8-jrj},
}

@article{merging_survey,
  author       = {Enneng Yang and
                  Li Shen and
                  Guibing Guo and
                  Xingwei Wang and
                  Xiaochun Cao and
                  Jie Zhang and
                  Dacheng Tao},
  title        = {Model Merging in LLMs, MLLMs, and Beyond: Methods, Theories, Applications
                  and Opportunities},
  journal      = {CoRR},
  volume       = {abs/2408.07666},
  year         = {2024},
  pages        = {},
  doi          = {10.48550/ARXIV.2408.07666},
}

@inproceedings{TTMM,
  author       = {Ryo Bertolissi and
                  Jonas H{\"{u}}botter and
                  Ido Hakimi and
                  Andreas Krause},
  title        = {Local Mixtures of Experts: Essentially Free Test-Time Training via
                  Model Merging},
  booktitle    = {The Second Conference on Language Modeling,
                  {CoLM} 2025},
  publisher    = {OpenReview.net},
  address      = {Montreal, Canada},
  year         = {2025},
  pages        = {},
  url          = {https://openreview.net/forum?id=X2RXpFA6Vh},
}


\appendix
\section{Experimental Setup}

\subsection{Dataset Statistics} \label{appendix:statistic}
Here we list the statistics of the model in Table~\ref{Table:dataset}.

\begin{table}[!htbp]
\caption{Dataset Statistics}
\centering
\begin{tabular}{c|cccccc}
    \hline
    Dataset & \#samples & \#field & \#feature & pos ratio \\
    \hline
    Criteo      & $4.6 \times 10^7$ & 39 & $6.8 \times 10^6$ & 0.2562 \\
    Avazu       & $4.0 \times 10^7$ & 24 & $4.4 \times 10^6$ & 0.1698 \\
    KDD12       & $1.5 \times 10^8$ & 11 & $6.0 \times 10^6$ & 0.0445 \\
    \hline
\end{tabular}
\begin{tablenotes}
\footnotesize
\item[1] Note: \textit{\#samples} refers to the total samples in the dataset, \textit{\#field} refers to the number of feature fields for original features, \textit{\#feature} refers to the number of unique feature values for original features, \textit{pos ratio} refers to the positive ratio. 
\end{tablenotes}
\label{Table:dataset}
\end{table}

\subsection{Hyper-parameters Setup}
\begin{table}[htbp]
\centering
\caption{Fixed hyperparameters for the original-size model.}
\label{tab:fixed_params}
\begin{tabular}{lc}
\toprule
Parameter & Value \\
\midrule
MLP dropout & 0.0 \\
Use batch normalization & False \\
MLP dimensions & [1024, 512, 256] \\
Cross layers & 3 \\
Interaction layers & 3 \\
LCB embedding dim & 16 \\
FMB embedding dim & 16 \\
FMB rank & 24 \\
FM MLP dimensions & [512, 512] \\
Number of L blocks & 3 \\
Expansion rate & 2 \\
Batch size & 4096 \\
\bottomrule
\end{tabular}
\end{table}

\begin{table}[htbp]
\centering
\caption{Varied configurations across datasets and original-size model variants.}
\label{tab:varied_params}
\setlength{\tabcolsep}{4pt}
\centering
\resizebox{0.48\textwidth}{!}{
\begin{tabular}{l|cccc}
\hline
Dataset & Model Variant & Latent Dim & LR & L2 Regularization \\
\hline
\multirow{11}{*}{Avazu(4.5M, 24)}
 & FNN & 16 & 0.0003 & 3e-6\\
 & IPNN & 16 & 0.0003 & 3e-6\\
 & DeepFM & 16 & 0.0003 & 1e-6\\
 & DCN & 16 & 0.0003 & 3e-6\\
 & DCNv2 & 16 & 0.003 & 0 \\
 & FinalMLP & 16 & 0.0003 & 3e-6\\
 & RankMixer & 24 & 0.0001 & 0 \\
 & Wukong & 16 & 0.0003 & 1e-6 \\
 & DCNv2+Wukong & 16 & 0.0003 & 1e-6 \\
 & DCNv2+RankMixer & 16/24 & 0.0001 & 0 \\
 & Wukong+RankMixer & 16/24 & 0.0001 & 0 \\
\hline
\multirow{11}{*}{Criteo(6.9M, 39)}
 & FNN & 16 & 0.0003 & 3e-6\\
 & IPNN & 16 & 0.0003 & 3e-6\\
 & DeepFM & 16 & 0.0003 & 3e-5\\
 & DCN & 16 & 0.0003 & 3e-6\\
 & DCNv2 & 16 & 0.001 & 3e-6 \\
 & FinalMLP & 16 & 0.0003 & 3e-6\\    
 & RankMixer & 39 & 0.001 & 1e-6 \\
 & Wukong & 16 & 0.0003 & 3e-6 \\
 & DCNv2+Wukong & 16 & 0.0003 & 3e-6 \\
 & DCNv2+RankMixer & 16/24 & 0.001 & 1e-6 \\
 & Wukong+RankMixer & 16/39 & 0.0003 & 3e-6 \\
\hline
\multirow{11}{*}{KDD12(6.1M, 11)}
 & FNN & 16 & 0.001 & 0\\
 & IPNN & 16 & 0.001 & 0\\
 & DeepFM & 16 & 0.001 & 0\\
 & DCN & 16 & 0.001 & 0\\
 & DCNv2 & 16 & 0.001 & 0 \\
  & FinalMLP & 16 & 0.001 & 0\\  
 & RankMixer & 33 & 0.0003 & 0 \\
 & Wukong & 16 & 0.0003 & 0 \\
 & DCNv2+Wukong & 16 & 0.0003 & 0 \\
 & DCNv2+RankMixer & 16/33 & 0.0001 & 0 \\
 & Wukong+RankMixer & 16/33 & 0.0003 & 0 \\
\hline
\end{tabular}}
\end{table}

\begin{table}[htbp]
\centering
\caption{Training and aggregation strategies for ensemble scales across datasets.}
\label{tab:ensemble_strategy}
\small
\setlength{\tabcolsep}{3pt}
\begin{tabular}{c|c|cccc}
\hline
\multirow{2}{*}{Dataset} & \multirow{2}{*}{Scale} & \multicolumn{4}{c}{Composition ($K \times [M // K]$)} \\
\cline{3-6}
 &  & DCNv2 & RankMixer & Wukong & Wukong+RankMixer \\
\hline
\multirow{7}{*}{\rotatebox{90}{Avazu}} 
 & 1x  & 1$\times$1 & 1$\times$1 & 1$\times$1 & 1$\times$1 \\
 & 2x  & 2$\times$1 & 2$\times$1 & 2$\times$1 & 2$\times$1 \\
 & 4x  & 4$\times$1 & 4$\times$1 & 2$\times$2 & 4$\times$1 \\
 & 8x  & 8$\times$1 & 8$\times$1 & 4$\times$2 & 8$\times$1 \\
 & 16x & 8$\times$2 & 16$\times$1 & 4$\times$4 & 16$\times$1 \\
 & 32x & 16$\times$2 & 16$\times$2 & 8$\times$4 & 16$\times$2 \\
 & 64x & 32$\times$2 & 16$\times$4 & 16$\times$4 & -- \\
\hline
\multirow{7}{*}{\rotatebox{90}{KDD12}} 
 & 1x  & 1$\times$1 & 1$\times$1 & 1$\times$1 & 1$\times$1 \\
 & 2x  & 1$\times$2 & 2$\times$1 & 1$\times$2 & 2$\times$1 \\
 & 4x  & 1$\times$4 & 2$\times$2 & 1$\times$4 & 2$\times$2 \\
 & 8x  & 1$\times$8 & 4$\times$2 & 1$\times$8 & 4$\times$2 \\
 & 16x & 1$\times$16 & 8$\times$2 & 1$\times$16 & 4$\times$4 \\
 & 32x & 1$\times$32 & 8$\times$4 & 1$\times$32 & 4$\times$8 \\
 & 64x & 1$\times$64 & 8$\times$8 & 1$\times$64 & 4$\times$16 \\
\hline
\end{tabular}

\end{table}

Table~\ref{tab:fixed_params} specifies the fixed hyperparameters for the original-size models, which serve as the baseline configuration prior to applying any scaling strategies.

Table~\ref{tab:varied_params} presents the dataset-specific hyperparameter configurations for the original-size model variants evaluated in this study. Each dataset is characterized by two dimensions shown in parentheses: the total number of features and the number of fields, respectively. For instance, the Avazu dataset contains 4.5 million features distributed across 24 fields.

Three critical hyperparameters were optimized for each model-dataset combination: latent dimension, learning rate (LR), and L2 regularization coefficient. The notation for latent dimensions requires clarification for hybrid models. Models connected with "+" (e.g., DCNv2+Wukong) represent joint training scenarios where both models share a single loss function and the same latent dimension. In contrast, the slash notation (e.g., 16/24) in models like DCNv2 + RankMixer indicates that the two component models use different latent dimensions, 16 for DCNv2 and 24 for RankMixer, when trained with separate loss functions.

The configurations show considerable variation across datasets, reflecting the diverse characteristics of each dataset and the specific architectural requirements of different models. This dataset-specific tuning proved essential for achieving optimal performance.


Table~\ref{tab:ensemble_strategy} documents the optimized training and aggregation strategies for ensemble scaling across different datasets and model variants. A key objective of our experimental design is to explore various training-aggregation configurations under fixed computational budgets to identify the most efficient scaling strategy. The notation ``$K \times [M // K]$'' describes two distinct aspects of the ensemble strategy under a fixed scale budget:
\begin{itemize}
    \item \textbf{$K$ (Number of training groups):} The number of models trained together, as shown in Algorithm~\ref{alg:tts} jointly via a single loss function.
    \item \textbf{$[M // K]$ (Model number in each group):} The number of such training ensembles whose predictions are aggregated for the final output, given Equation~\ref{eq:average}.
\end{itemize}
For all dataset-model combinations not listed in Table~\ref{tab:ensemble_strategy}, all models are independently trained without any group training.

\subsection{Parameter Scaling Dimension Strategies on Given Models}
\label{appendix:experiment_scaling_dim}
\begin{table}[htbp]
\centering
\caption{Hyperparameters for Wukong and Rankmixer.}
\label{tab:param_scaling_hyperparams}

\vspace{-10pt}
\footnotesize
\setlength{\tabcolsep}{2.5pt}
\textbf{(a) Wukong model scaling strategies (Wukong-specific parameters)}
\vspace{0.3em}

\begin{tabular}{c|c|cccccc}
\hline
\multirow{3}{*}{Strategy} & \multirow{3}{*}{Scale} & \multicolumn{6}{c}{Wukong-Specific Hyperparameters} \\
\cline{3-8}
 & & MLP & Interaction & LCB & FMB & FMB & FM MLP \\
 & & Dims & Layers & Embed. & Embed. & Rank & Dims \\
\hline
\multirow{4}{*}{I} 
 & 2x  & [512, 512] & 6 & 16 & 16 & 24 & [512, 512] \\
 & 4x  & [512, 512] & 12 & 16 & 16 & 24 & [512, 512] \\
 & 8x  & [512, 512] & 24 & 16 & 16 & 24 & [512, 512] \\
 & 16x & [512, 512] & 48 & 16 & 16 & 24 & [512, 512] \\
\hline
\multirow{4}{*}{K\_Nf\_Nl}
 & 2x  & [512, 512] & 3 & 32 & 32 & 48 & [512, 512] \\
 & 4x  & [512, 512] & 3 & 64 & 64 & 96 & [512, 512] \\
 & 8x  & [512, 512] & 3 & 128 & 128 & 192 & [512, 512] \\
 & 16x & [512, 512] & 3 & 256 & 256 & 384 & [512, 512] \\
\hline
\end{tabular}

\vspace{1em}

\footnotesize
\setlength{\tabcolsep}{3pt}
\textbf{(b) RankMixer model scaling strategies (RankMixer-specific parameters)}
\vspace{0.3em}

\begin{tabular}{c|c|ccc}
\hline
\multirow{2}{*}{Strategy} & \multirow{2}{*}{Scale} & \multicolumn{3}{c}{RankMixer-Specific Hyperparameters} \\
\cline{3-5}
 &  & Latent Dim & Number of L Blocks & Expansion Rate \\
\hline
\multirow{3}{*}{D} 
 & 2x  & 48 & 3 & 2 \\
 & 4x  & 96 & 3 & 2 \\
 & 8x  & 192 & 3 & 2 \\
\hline
\multirow{3}{*}{D\_L} 
 & 4x  & 48 & 6 & 2 \\
 & 8x  & 96 & 6 & 2 \\
 & 16x & 192 & 6 & 2 \\
\hline
\multirow{3}{*}{L} 
 & 2x  & 24 & 6 & 2 \\
 & 4x  & 24 & 12 & 2 \\
 & 8x  & 24 & 24 & 2 \\
\hline
\multirow{1}{*}{T} 
 & 2x  & 24 & 3 & 4 \\
\hline
\end{tabular}
\vspace{1em}

\textbf{(c) DCNv2 model scaling strategies (DCNv2-specific parameters)}
\vspace{0.3em}

\begin{tabular}{c|c|cc}
\hline
\multirow{2}{*}{Strategy} & \multirow{2}{*}{Scale} & \multicolumn{2}{c}{DCNv2-Specific Hyperparameters} \\
\cline{3-4}
 &  & MLP Dims & Cross Layers \\
\hline
\multirow{1}{*}{W}
 & 2x  & [2048, 1024, 512] & 3 \\
\hline
\multirow{4}{*}{D} 
 & 2x  & [1024, 1024, 512, 512, 256, 256] & 6 \\
 & 4x  & [1024$\times$4, 512$\times$4, 256$\times$4] & 12 \\
 & 8x  & [1024$\times$8, 512$\times$8, 256$\times$8] & 24 \\
 & 16x & [1024$\times$16, 512$\times$16, 256$\times$16] & 48 \\
\hline
\multirow{3}{*}{O} 
 & 4x  & [2048, 1024$\times$3, 512, 256] & 6 \\
 & 8x  & [2048$\times$2, 1024$\times$3, 512$\times$3, 256] & 9 \\
 & 16x & [4096, 2048$\times$3, 1024$\times$3, 512$\times$4, 256] & 12 \\
\hline
\end{tabular}

\vspace{1em}

\footnotesize
\setlength{\tabcolsep}{3pt}
\textbf{(d) FinalMLP model scaling strategies (FinalMLP-specific parameters)}
\vspace{0.3em}

\begin{tabular}{c|c|c}
\hline
\multirow{2}{*}{Strategy} & \multirow{2}{*}{Scale} & FinalMLP-Specific Hyperparameter \\
\cline{3-3}
 &  & MLP Dims \\
\hline
\multirow{1}{*}{W}
 & 2x  & [2048, 1024, 512] \\
\hline
\multirow{4}{*}{D} 
 & 2x  & [1024, 1024, 512, 512, 256, 256] \\
 & 4x  & [1024$\times$4, 512$\times$4, 256$\times$4] \\
 & 8x  & [1024$\times$8, 512$\times$8, 256$\times$8] \\
 & 16x & [1024$\times$16, 512$\times$16, 256$\times$16] \\
\hline
\multirow{3}{*}{O} 
 & 4x  & [2048, 1024$\times$3, 512, 256] \\
 & 8x  & [2048$\times$2, 1024$\times$3, 512$\times$3, 256] \\
 & 16x & [4096, 2048$\times$3, 1024$\times$3, 512$\times$4, 256] \\
\hline
\end{tabular}
\vspace{-10pt}
\end{table}

Table~\ref{tab:param_scaling_hyperparams} presents model-specific parameters for each architecture: Wukong parameters include MLP dimensions, interaction layers, LCB/FMB embedding dimensions, FMB rank, and FM MLP dimensions; RankMixer parameters encompass latent dimensions, number of L blocks, and expansion rate; DCNv2 parameters consist of MLP dimensions and cross layers; FinalMLP parameters comprise MLP dimensions only.

The scaling strategies are defined as follows:
\begin{itemize}
    \item \textbf{I}: Increase the number of interaction layers in the Wukong model
    \item \textbf{K\_Nf\_Nl}: Scale embedding dimensions (including Rank, LCB, and FMB components) in the Wukong model
    \item \textbf{W}: Apply width scaling to multi-layer perceptron dimensions in DCNv2 and FinalMLP models
    \item \textbf{D}: Scale depth by repeating MLP layers in DCNv2 and FinalMLP models
    \item \textbf{O}: Apply optimized scaling with wider layers in DCNv2 and FinalMLP models
    \item \textbf{L}: Scale the number of L blocks exclusively in the Rankmixer model
    \item \textbf{D\_L}: Jointly scale both latent dimensions and the number of L blocks in the Rankmixer model
    \item \textbf{T}: Scale the expansion rate in the Rankmixer model
\end{itemize}
\section{Extended Evaluations}

\subsection{Further Scalability on Selected Components}
\label{appendix:extended}
\begin{table}[!htbp]
    \renewcommand\arraystretch{1.2}
    \centering
    \caption{Extended Performance Comparison with Higher Randomness (up to 64×) and Ensemble Combinations.}
    \label{tab:main_homo_extended}
    \vspace{-10pt}
    \resizebox{0.48\textwidth}{!}{
    \begin{tabular}{c|c|cc|cc|cc}
        \hline
                                                & \multirow{2}{*}{Method} & \multicolumn{2}{c|}{DCNv2} & \multicolumn{2}{c|}{RankMixer} & \multicolumn{2}{c}{Wukong} \\
        \cline{3-8}                            &                         & AUC$\uparrow$              & Logloss$\downarrow$                 & AUC$\uparrow$                  & Logloss$\downarrow$      & AUC$\uparrow$             & Logloss$\downarrow$                \\
        \hline
        \multirow{7}{*}{\rotatebox{90}{Avazu}}  & 1$\times$               & 0.7946                     & 0.3711                         & 0.7952                         & 0.3706              & 0.7954                    & 0.3704                        \\
                                                & 2$\times$               & 0.7953                     & 0.3706                         & 0.7962                         & 0.3699              & 0.7972                    & 0.3694                        \\
                                                & 4$\times$               & 0.7955                     & 0.3704                         & 0.7972                         & 0.3694              & 0.7986                    & 0.3685                        \\
                                                & 8$\times$               & 0.7958                     & 0.3703                         & 0.7981                         & 0.3689              & 0.7994                    & 0.3681                        \\
                                                & 16$\times$              & 0.7959                     & 0.3702                         & 0.7988                         & 0.3684              & 0.8001                    & 0.3677                        \\
                                                & 32$\times$              & 0.7960                     & 0.3701                         & 0.7994                         & 0.3680              & {0.8006}           & {0.3674}                        \\
                                                & 64$\times$              & 0.7960                     & 0.3701                         & 0.7997                         & 0.3679              & 0.8003                    & 0.3676                        \\
        \hline
        \multirow{7}{*}{\rotatebox{90}{Criteo}} & 1$\times$               & 0.8116                     & 0.4398                         & 0.8115                         & 0.4400              & 0.8103                    & 0.4411                        \\
                                                & 2$\times$               & 0.8124                     & 0.4391                         & 0.8124                         & 0.4392              & 0.8121                    & 0.4395                        \\
                                                & 4$\times$               & 0.8129                     & 0.4386                         & 0.8133                         & 0.4383              & 0.8131                    & 0.4387                        \\
                                                & 8$\times$               & 0.8131                     & 0.4384                         & 0.8139                         & 0.4378              & 0.8135                    & 0.4383                        \\
                                                & 16$\times$              & 0.8132                     & 0.4384                         & 0.8142                         & 0.4375              & 0.8137                    & 0.4381                        \\
                                                & 32$\times$              & 0.8133                     & 0.4383                         & 0.8142                         & 0.4375              & 0.8139                    & 0.4380                        \\
                                                & 64$\times$              & 0.8133                     & 0.4383                         & 0.8143                         & 0.4374              & 0.8139                    & 0.4379                        \\
        \hline
        \multirow{7}{*}{\rotatebox{90}{KDD12}}  & 1$\times$               & 0.8114                     & 0.1499                         & 0.8124                         & 0.1496              & 0.8117                    & 0.1498                        \\
                                                & 2$\times$               & 0.8119                     & 0.1497                         & 0.8133                         & 0.1494              & 0.8128                    & 0.1496                        \\
                                                & 4$\times$               & 0.8122                     & 0.1497                         & 0.8139                         & 0.1493              & 0.8133                    & 0.1495                        \\
                                                & 8$\times$               & 0.8123                     & 0.1496                         & 0.8144                         & 0.1492              & 0.8135                    & 0.1494                        \\
                                                & 16$\times$              & 0.8124                     & 0.1496                         & 0.8148                         & 0.1491              & 0.8137                    & 0.1494                        \\
                                                & 32$\times$              & 0.8124                     & 0.1496                         & 0.8151                         & 0.1490              & 0.8138                    & 0.1494                        \\
                                                & 64$\times$              & 0.8124                     & 0.1496                         & 0.8152                         & 0.1489              & 0.8138                    & 0.1494                        \\
        \hline
    \end{tabular}
    }

    \vspace{-10pt}
\end{table}
In this section, we further scale selected models given the setup in Table~\ref{table:main_homo} and Section~\ref{sec:experiment_main_homo}. Specifically, we scale up the homogenous models by up to 64$\times$ based on DCNv2, Rankmixer, and Wukong. The results are shown in Table~\ref{tab:main_homo_extended}. We can witness a performance plateau after 16$\times$.

\subsection{Hyperparameter Study}
\label{appendix:hyperparam}
In this section, we conduct a hyperparameter ablation on two combinations: Wukong on Avazu and Rankmixer on KDD12 due to their special configurations in Table~\ref{tab:ensemble_strategy}. Based on the results in Table~\ref{tab:config_ablation}, we can observe that group training has a non-trivial effect on the final outcomes.

\begin{table}[!htbp]
\centering
\caption{Performance under different configurations.}
\label{tab:config_ablation}
\vspace{-10pt}
\begin{tabular}{l|cc|cc}
\hline
\multirow{2}{*}{Config}& \multicolumn{2}{c|}{AVAZU (Wukong)} & \multicolumn{2}{c}{KDD (RankMixer)}\\
\cline{2-5}
    & AUC$\uparrow$ & LogLoss$\downarrow$ & AUC$\uparrow$ & LogLoss$\downarrow$\\
\hline
1$\times$16   & 0.7984 & 0.3686 & 0.8136 & 0.1493 \\
2$\times$8    & 0.7997 & 0.3678 & 0.8143 & 0.1492 \\
4$\times$4    & \textbf{0.8001} & \textbf{0.3677} & 0.8146 & 0.1491 \\
8$\times$2    & 0.7999 & 0.3678 & \textbf{0.8148} & \textbf{0.1491} \\
16$\times$1   & 0.7982 & 0.3690 & O.O.M. & O.O.M. \\
\hline
\end{tabular}
\end{table}

\end{document}